\documentclass[amsmath, aps]{revtex4}
\usepackage[pdftex]{graphicx}
\usepackage{times}
\usepackage{epsfig}
\usepackage{amssymb}
\usepackage{multirow}

\usepackage{amsmath}
\begin{document}
\title{Theory of the multiphoton cascade transitions with two photon links: comparison of Quantum Electrodynamical and Quantum Mechanical approaches}
\author{T. Zalialiutdinov$^{1}$, Yu. Baukina$^{1}$, D. Solovyev$^1$ and  L. Labzowsky$^{1,2}$}

\affiliation{ 
$^1$ V. A. Fock Institute of Physics, St. Petersburg
State University, Petrodvorets, Oulianovskaya 1, 198504,
St. Petersburg, Russia
\\
$^2$  Petersburg Nuclear Physics Institute, 188300, Gatchina, St.
Petersburg, Russia}
\begin{abstract}
QED theory of multiphoton cascade transitions in atoms and ions is developed. In particular the $ 3s\rightarrow1s+2\gamma $, $ 4s\rightarrow1s+2\gamma $ and $ 3p\rightarrow1s+3\gamma $ processes are considered. Two different approaches (Quantum Electrodynamical and phenomenological Quantum Mechanical) are applied.
\end{abstract}
\maketitle

\section{Introduction}

The theory of the multiphoton transitions in atoms on the basis of Quantum Mechanics (QM) started with the work by G\"{o}ppert-Mayer \cite{1}. The first evaluation of the two-photon decay rate $ 2s\rightarrow 1s+2\gamma(E1) $ in hydrogen was performed by Breit and Teller \cite{2} (see correction to this work in \cite{3}). Accurate nonrelativistic evaluation of $ 2s\rightarrow1s+2\gamma(E1) $ transition rate in hydrogen was performed in \cite{4}. The first fully relativistic calculation of this transition applicable also to the highly charged H-like ions (H-like HCI) was given in \cite{5} and later in \cite{6}, \cite{7} and \cite{8}. Quantum Electrodynamical (QED) corrections to the two-photon decay of H-like ions were studied in \cite{9}, \cite{10}.  The calculations of different two-photon transition rates in hydrogen atom and H-like ions were given in \cite{11}-\cite{13} in the nonrelativistic approximation and in a fully relativistic approach in \cite{14}-\cite{18}.

Accurate evaluations of the two-photon transition rates in hydrogen are important for several reasons. First, during the last decades the two-photon $ 2s-1s$ transition frequency measurements reached exclusively high level of accuracy \cite{19}-\cite{21} and the accurate measurements of the two-photon $3s-1s$, $3d-1s$ frequency transitions were also reported \cite{22}, \cite{23}. Theoretical studies of the accuracy limits for these measurements require the knowledge of the corresponding transition rates \cite{24}. Second, the two-photon transition rates in hydrogen are important for astrophysics, since these processes were involved in the formation of the Cosmic Microwave Background (CMB) during the epoch of the cosmological hydrogen recombination. Recent accurate measurements of the properties of CMB also require the accurate knowledge of the two-photon $2s-1s$, $3s-1s$, $3d-1s$ etc. transition rates in hydrogen. For this purpose the two-photon transition rates in hydrogen were evaluated in \cite{25}-\citep{28}. 

There is an essential difference in the description of the two-photon $2s-1s$ and the two-photon transitions to the ground state from the higher levels.This difference is due to the presence of the cascade contributions which are absent only in $2s-1s$ transition. The cascade problem in the two-photon transitions in two-electron HCI was first discussed in \cite{29}. The same problem was considered later in \cite{30}. In \cite{31} (see also \cite{32}) a general QED approach was developed which allows for the description of cascade transitions. In the theory of the cascade transition rates an important question is the regularization of the singular cascade contributions to the total two-photon transition rate. This regularization is performed by the introduction of the level widths in the energy denominators which otherwise turn to zero for the photon frequencies, corresponding to the cascade resonances. In the QM phenomenological approach only the width of the intermediate state is usually employed for regularization \cite{12}, \cite{13}, \cite{27}. We call this approach phenomenological since the width is introduced as an phenomenological parameter. In the QED approach \cite{31}, \cite{32} based on the Low derivation of the Lorentz profile from QED \cite{33}, the regularization is performed via the introduction of the sum of two widths, for the initial and intermediate states. Note, that in \cite{13} the possibility of using the sum of the widths was briefly mentioned. In \cite{34}-\cite{36} the QED approach for the description of the multiphoton transitions with cascades was applied to the decays of $3s$, $3p$, $3d$ and $4s$ states in the hydrogen atom.

In the present paper we compare both methods of regularization. A detailed derivation of the QED regularization approach is given. We also demonstrate that QM phenomenological approach being properly used gives the same result as QED, i.e. with the sum of two widths. The importance of the correct regularization will be also illustrated.

It is convenient to describe the regularizations of cascade contributions using the simplest example: the two-photon transition $ 3s\rightarrow1s+2\gamma $. In $ 3s\rightarrow 1s+2\gamma $ transition a single cascade $ 3s\rightarrow2p+\gamma\rightarrow1s+2\gamma $ should be taken into account.

A total transition rate $ W^{2\gamma}_{3s-1s} $ can be written as
\begin{eqnarray}
\label{1}
W^{2\gamma}_{3s-1s}=\frac{1}{2}\int\limits^{\omega_0}_0dW^{2\gamma}_{3s-1s}(\omega)\;,
\end{eqnarray}
where $ dW^{2\gamma}_{3s-1s}(\omega) $ is the differential transition rate, $ \omega $ is the frequency of one of the emitted photons, $ \omega_0=E_{3s}-E_{1s} $. The differential transition rate $dW^{2\gamma}_{3s-1s}(\omega)$ consists of three terms: cascade contribution, "pure" two-photon contribution and the interference contribution:
\begin{eqnarray}
\label{2}
dW^{2\gamma}_{3s-1s}=dW^{2\gamma(cascade)}_{3s-1s}+dW^{2\gamma(pure)}_{3s-1s}+dW^{2\gamma(interference)}_{3s-1s}\;.
\end{eqnarray}
A possibility of separation of the cascade contribution from the "pure"\;two-photon contribution in Eq. (\ref{2}) was discussed in several papers: in \cite{31} for the two-electron HCI and in \cite{37}-\cite{39}, \cite{34} for the two-photon transitions in hydrogen. An ambiguity of this separation was demonstrated in \cite{34}. The cascade contribution can be presented as the sum of the contributions of two cascade links (resonances):
\begin{eqnarray}
\label{3}
dW^{2\gamma(cascade)}=dW^{2\gamma(resonance\;1)}_{3s-2p-1s}+dW^{2\gamma(resonance\;2)}_{3s-2p-1s}\;,
\end{eqnarray}
where two resonant frequencies are: $ \omega^{res1}=E_{3s}-E_{2p} $ and $ \omega^{res2}=E_{2p}-E_{1s}$. The corresponding resonance contributions were presented in \cite{34} on the basis of the QED approach 
\begin{eqnarray}
\label{4}
dW^{2\gamma(resonance\;1)}_{3s-2p-1s}=\frac{\Gamma_{3s}+\Gamma_{2p}}{\Gamma_{2p}}\frac{W^{1\gamma}_{3s-2p}(\omega^{res1})W^{1\gamma}_{2p-1s}(\omega^{res2})d\omega}{(\omega-\omega^{res1})^2+\frac{1}{4}(\Gamma_{3s}+\Gamma_{2p})^2}\;,
\end{eqnarray}

\begin{eqnarray}
\label{5}
dW^{2\gamma(resonance\;2)}_{3s-2p-1s}=\frac{W^{1\gamma}_{3s-2p}(\omega^{res1})W^{1\gamma}_{2p-1s}(\omega^{res2})d\omega}{(\omega-\omega^{res2})^2+\frac{1}{4}\Gamma_{2p}^2}\;.
\end{eqnarray}

Note that the factor $ \frac{\Gamma_{3s}+\Gamma_{2p}}{\Gamma_{2p}} $ in Eq. (\ref{4}) was lost in \cite{34} though this led to relatively small numerical error since $ \Gamma_{3s}\approx 0.01 \Gamma_{2p} $.

Here $ \Gamma_{3s} $, $ \Gamma_{2p} $ are the total widths of the levels $ 3s $, $ 2p $ and $ W^{1\gamma}_{3s-2p} $, $ W^{1\gamma}_{2p-1s} $ are the one-photon transition rates. In the nonrelativistic limit $ \Gamma_{3s}=W^{1\gamma}_{3s-2p} $,  $ \Gamma_{2p}=W^{1\gamma}_{2p-1s}$. Then, integrating Eqs. (\ref{4}), (\ref{5}) over $ \omega $ and taking into account Eq. (\ref{1}) we find

\begin{eqnarray}
\label{6}
\frac{1}{2}\int\limits^{\omega_0}_0dW^{2\gamma(resonance\;1)}_{3s-2p-1s}=\frac{1}{2}W^{1\gamma}_{3s-2p}=\frac{1}{2}\Gamma_{3s}\;,
\end{eqnarray}

\begin{eqnarray}
\label{7}
\frac{1}{2}\int\limits^{\omega_0}_0dW^{2\gamma(resonance\;2)}_{3s-2p-1s}=\frac{1}{2}W^{1\gamma}_{3s-2p}=\frac{1}{2}\Gamma_{3s}\;.
\end{eqnarray}
Hence,
\begin{eqnarray}
\label{7a}
 W^{2\gamma(cascade)}_{3s-1s}=\Gamma_{3s} 
\end{eqnarray}
and
\begin{eqnarray}
\label{8}
W^{2\gamma}_{3s-1s}=\Gamma_{3s}+\frac{1}{2}\int\limits^{\omega_0}_0[dW^{2\gamma(pure)}_{3s-1s}+dW^{2\gamma(interference)}_{3s-1s}]\;.
\end{eqnarray}
Within the QM approach the regularization was usually performed by introducing the widths for the intermediate $ np $ states, i.e. replacing the energy $ E_{np} $ by $ E_{np}-\frac{i}{2}\Gamma_{np} $. In case of $ 3s-2p-1s $ cascade $ \Gamma_{np}=\Gamma_{2p} $. Therefore the contribution of the resonance 1 instead of Eq. (\ref{4}) looked like
\begin{eqnarray}
\label{9}
dW^{2\gamma(resonance\;1)}_{3s-2p-1s}=\frac{W^{1\gamma}_{3s-2p}(\omega^{res1})W^{1\gamma}_{2p-1s}(\omega^{res2})d\omega}{(\omega-\omega^{res1})^2+\frac{1}{4}\Gamma_{2p}^2}\;,
\end{eqnarray}
while the contribution of $ dW^{2\gamma(resonance\;2)}_{3s-2p-1s} $ remained the same as in Eq. (\ref{5}). 

We have to stress that the insertion of Eq. (\ref{9}) in the integral Eq. (\ref{1}) gives exactly the same result Eq. (\ref{6}) as the insertion  of Eq. (\ref{4}). One could think that both methods of regularizations are equivalent. However, this is not the case for two reasons. First, this equivalence for the cascade contributions is approximate. Eqs. (\ref{6}), (\ref{7}) are valid up to the small corrections of the order $ \Gamma_{3s}/\omega_{0} $. With the same accuracy holds the mentioned equivalence. Second, in the different applications the frequency distributions for the two-photon decays are converted with some other functions. This also violates the equivalence mentioned above. 

In the second part of the paper we address the same problem of the cascade regularization within the QM approach following mainly \cite{40}. In \cite{40} the derivation is given for one-photon transitions between two excited atomic states, which corresponds to one "link"\;of the two-photon cascade transition. We extend this derivation to the full two-photon transition rate, including cascade, "pure" two-photon and interference contributions. We demonstrate that with the proper treatment of QM approach we arrive again at the expression Eq. (\ref{4}) (and not Eq. (\ref{9})) for the cascade contribution in case of the $ 3s\rightarrow 1s +2\gamma $ transition. The QM approach is then generalized also to the 3-photon and 4-photon transitions. In all cases the QM approach gives exactly the same results as the QED theory.

The paper is organized as follows. In section II we start with the QED derivation of the Lorentz profile for the one-photon transition from the excited state to the ground state. This derivation repeats shortly the derivations in \cite{31}, \cite{32} but is necessary to introduce the basic formulas and notations. As an example the Lyman-alpha transition $ 2p-1s $ is considered. In section III the two-photon transition rate to the ground state from the $ ns $-state in the presence of cascades is described in general. In section IV the regularization of the two-photon cascade transition $ 3s-1s $ is analysed. In section V the non-cascade contributions to the $ 3s-1s$  two photon transition rate are described. In section VI the same derivations are made for the two-photon $ 4s-1s $ transition: there is an important difference between $ 4s-1s $ and $ 3s-1s $ two-photon transitions due to the existence of several cascade channels in case of $ 4s-1s $. In section VII the 3-photon transitions are analysed in the "two-photon approximation" with 3-photon decay $ 3p-1s $ as an example. In the section VIII the QM approach is applied to the two- and three-photon transitions. In the section IX the importance of the correct regularization is demonstrated. Concluding remarks are presented in section X. 

\section{One-photon transition to the ground state}

The full QED description of any process in an atom should start with the ground state and end up with the ground state too, i.e. the excitation of the decaying state should be always included. For the resonant processes, e.g. for the resonant photon scattering the absorption part of the process can be well separated from the emission part, so that the description of the decay process independent on the excitation becomes possible. In this way the theory of the multiphoton processes in atoms was developed in  \cite{31}, \cite{32}.

Having in mind the recombination processes in hydrogen atom we consider first the resonance photon scattering on the ground $ 1s $ state with resonances corresponding to the $ np $ states. In our derivations we will fully neglect the photons other than $ E1 $ which is reasonable for the neutral hydrogen. It is important to stress that we consider the free atoms which are excited by the photons released by the source which line widths is comparable (or larger) then the natural line widths of the resonance atomic state. Thus we exclude the special cases of the excitation by the laser with the narrow bandwidths or something equivalent. Our condition (broad source width) should correspond the cosmological recombination situation when every atom is excited by the photons emitted by another atom. The Feynman graph corresponding to the resonant photon scattering is depicted in Fig. 1a.

The $ S $-matrix element, corresponding to Fig. 1a, i.e. second-order scattering process, looks like
\begin{eqnarray}
\label{11}
S_{1s}^{(2)sc}=(-ie)^2\int d^4x_1d^4x_2\overline{\psi}_{1s}(x_1)\gamma_{\mu_1}A_{\mu_1}^{*(\vec{k}_f\vec{e}_f)}(x_1)S(x_1,x_2)\gamma_{\mu_2}A_{\mu_2}^{(\vec{k}_i\vec{e}_i)}(x_2)\psi_{1s}(x_2)\;,
\end{eqnarray}
where
\begin{eqnarray}
\label{12}
\psi_A(x)=\psi_A(\vec{r})e^{-iE_At}\;,
\end{eqnarray}
$ \psi_A(\vec{r}) $ is the solution of the Dirac equation for the atomic electron, $ E_A $ is the Dirac energy, $ \overline{\psi}_A=\psi^+_A\gamma_0 $ is the Dirac conjugated wave function, $ \gamma_{\mu}\equiv(\gamma_0,\vec{\gamma})$ are the Dirac matrices and $ x\equiv(\vec{r},t) $ is the space-time coordinate. The photon field or the photon wave function $ A_{\mu}(x) $ looks like
\begin{eqnarray}
\label{13}
A_{\mu}^{(\vec{k},\vec{e})}=\sqrt{\frac{2\pi}{\omega}}e_{\mu}e^{i(\vec{k}\vec{r}-\omega t)}=\sqrt{\frac{2\pi}{\omega}}e^{-i\omega t}A_{\mu}^{(\vec{k},\vec{e})}(\vec{r})\;,
\end{eqnarray}
where $ e_{\mu} $ are the components of the photon polarization four-vector ($ \vec{e} $ is 3-dimensional polarization vector for real photons), $ k\equiv(\vec{k},\omega) $ is the photon momentum four-vector, $ \vec{k} $ is the wave vector, $ \omega=|\vec{k}| $ is the photon frequency. Eq. (\ref{13}) corresponds to the absorbed photon and  $ A^{*(\vec{k},\vec{e})}_{\mu} $ corresponds to the emitted photon. Finally, the electron propagator for the bound electron it is convenient to present in the form of the eigenmode decomposition with respect to one-electron eigenstates \cite{40}, \cite{41}
\begin{eqnarray}
\label{14}
S(x_1,x_2)=\frac{1}{2\pi i}\int\limits^{\infty}_{-\infty}d\omega e^{-i\omega(t_1-t_2)}\sum\limits_{n}\frac{\psi_n(\vec{r}_1)\overline{\psi}_n(\vec{r}_2)}{\omega-E_n(1-i0)}\;.
\end{eqnarray}
Insertion of the expressions (\ref{12})-(\ref{14}) into Eq. (\ref{11}) and integration over time and frequency variables leads to
\begin{eqnarray}
\label{15}
S_{1s}^{(2)sc}=-2\pi i\delta(\omega_f-\omega_i)e^2\sum\limits_{n}\frac{(\gamma_{\mu}A^{*(\vec{k}_f,\vec{e}_f)}_{\mu})_{1sn}(\gamma_{\mu}A^{(\vec{k}_i,\vec{e}_i)}_{\mu})_{n1s}}{\omega_f+E_{1s}-E_n}\;.
\end{eqnarray}
The amplitude $ U $ of the elastic photon scattering is related to the $ S $-matrix element via \cite{41}
\begin{eqnarray}
\label{16}
S=-2\pi i\delta(\omega_f-\omega_i)U\;.
\end{eqnarray}
Accordingly, we will obtain the scattering amplitude 
\begin{eqnarray}
\label{17}
U^{(2)sc}_{1s}=e^2\sum\limits_{n}\frac{(\gamma_{\mu}A^{*(\vec{k}_f,\vec{e}_f)}_{\mu})_{1sn}(\gamma_{\mu}A^{(\vec{k}_i,\vec{e}_i)}_{\mu})_{n1s}}{\omega_f+E_{1s}-E_n}\;,
\end{eqnarray}
where the energy conservation law implies that $ |\vec{k_f}|=|\vec{k_i}| $.

For the resonant scattering process the photon frequency $ \omega_i=\omega_f $ is close to the energy difference between two atomic levels. In case of $ np $ resonance $ \omega_i\simeq E_{np}-E_{1s} $. Accordingly we have to retain only one term in the sum over $ n $ in Eq. (\ref{17})
\begin{eqnarray}
\label{18}
U^{(2)sc}_{1s(np)}=e^2\frac{(\gamma_{\mu}A^{*(\vec{k}_f,\vec{e}_f)}_{\mu})_{1snp}(\gamma_{\mu}A^{(\vec{k}_i,\vec{e}_i)}_{\mu})_{np1s}}{\omega_f+E_{1s}-E_{np}}\;.
\end{eqnarray}

Eq. (\ref{18}) reveals that in the resonance approximation the scattering amplitude is factorized into an emission and absorption parts. The energy denominator should be attached to the emission or absorption part depending on what we want to describe: emission or absorption process. In particular, the first-order emission amplitude can be expressed as
\begin{eqnarray}
\label{19}
U^{em}_{np1s}=e\frac{(\gamma_{\mu}A^{*(\vec{k}_f,\vec{e}_f)}_{\mu})_{1snp}}{\omega_f+E_{1s}-E_{np}}\;.
\end{eqnarray}

The nonresonant corrections to the resonance approximation, first introduced  in \cite{33} were recently investigated in \cite{42}, \cite{43}, \cite{34}. The role of these corrections appeared to be negligible in most cases. These corrections arise when one takes into account the terms other than the resonant one in sum over $ n $ in Eq. (\ref{17}). The same concerns nonresonant contribution to the scattering amplitude which arises when we interchange the position of the photon lines in Fig. 1, i.e. when the emission of the photon occurs prior to the absorption.

The energy conservation law which follows from Eq. (\ref{16}) reads
\begin{eqnarray}
\label{20a}
\omega_i=\omega_f\;.
\end{eqnarray}
The resonance condition one can write in the form:
\begin{eqnarray}
\label{20b}
|\omega_i-E_{np}+E_{1s}|=|\omega_f-E_{np}+E_{1s}|\leqslant \Gamma_{np}\;.
\end{eqnarray}
In cases, when we can neglect $ \Gamma_{np} $ in Eq. (\ref{20b}) this equation takes the form of the energy conservations law
\begin{eqnarray}
\label{20c}
\omega_f=E_{np}-E_{1s}\;.
\end{eqnarray}
In particular we can use Eq. (\ref{20c}) in the numerator of Eq. (\ref{19}) but not in its denominator.

To derive the Lorentz profile for the emission process we follow the Low procedure \cite{33}, i.e. insert infinite number of the self-energy corrections in the resonance approximation into the electron propagator in Fig. 1a. The first term of the corresponding Feynman graph sequence is depicted in Fig. 1b. Employing the photon propagator in the Feynman gauge in the form
\begin{eqnarray}
\label{20}
D_{\mu_1 \mu_2}(x_1-x_2)=\frac{1}{2\pi i}\int\limits^{\infty}_{-\infty}d\Omega I_{\mu_1 \mu_2}(|\Omega|, r_{12})e^{-i\Omega(t_1-t_2)}\;,
\end{eqnarray}
\begin{eqnarray}
\label{21}
I_{\mu_1 \mu_2}=\frac{\delta_{\mu_1 \mu_2}}{r_{12}}e^{i|\Omega| r_{12}}\;,
\end{eqnarray}
where $ x\equiv(\vec{r}, t) $, $ r_{12}=|\vec{r}_1-\vec{r}_2| $ and defining the matrix element of the electron self-energy operator as \cite{44}
\begin{eqnarray}
\label{22}
(\widehat{\Sigma}(\xi))_{AB}=\frac{e^2}{2\pi i}\sum\limits_{n}\int d\Omega\frac{(\gamma_{\mu_1}\gamma_{\mu_2}I_{\mu_1\mu_2}(|\Omega|,r_{12}))_{AnnB}}{\xi-\Omega-E_n(1-i0)}\;,
\end{eqnarray}
 we obtain the following expression for the correction to the scattering amplitude \cite{32}:
\begin{eqnarray}
\label{23}
U^{(4)sc.}_{1s}=e^2\sum\limits_{n_1 n_2}\frac{(\gamma_{\mu}A^{*(\vec{k}_f,\vec{e}_f)}_{\mu})_{1sn_1}(\widehat{\Sigma}(\omega+E_{1s}))_{n_1n_2}(\gamma_{\mu}A^{(\vec{k}_i,\vec{e}_i)}_{\mu})_{n_21s}}{(\omega_f+E_{1s}-E_{n_1})(\omega+E_{1s}-E_{n_2})}\;.
\end{eqnarray}
The resonance approximation implies $ n_1=n_2=np $. Then taking into account Eq. (\ref{18}) we can write
\begin{eqnarray}
\label{24}
U^{(4)sc.}_{1s(np)}=U^{(2)sc.}_{1s(np)}\frac{(\widehat{\Sigma}(\omega_f+E_{1s}))_{np,np}}{\omega_f+E_{1s}-E_{np}}\;.
\end{eqnarray}
Repeating these insertions in the resonance approximation leads to a geometric progression. Summation of this progression yields
\begin{eqnarray}
\label{25}
U^{(4)sc.}_{1s(np)}=e^2\frac{\gamma_{\mu}A^{*(\vec{k}_f,\vec{e}_f)}_{\mu})_{1snp}(\gamma_{\mu}A^{(\vec{k}_i,\vec{e}_i)}_{\mu})_{np1s}}{(\omega_f+E_{1s}-E_{2p}-(\widehat{\Sigma}(\omega_f+E_{1s}))_{np,np}}\;.
\end{eqnarray}
The emission amplitude looks like
\begin{eqnarray}
\label{26}
U^{em}_{np 1s}=e\frac{(\gamma_{\mu}A^{*(\vec{k}_f\vec{e}_f)}_{\mu})_{1snp}}{\omega_f+E_{1s}-E_{np}-(\widehat{\Sigma}(\omega_f+E_{1s}))_{np,np}}\;.
\end{eqnarray}
The operator $ \widehat{\Sigma}(\omega_f+E_{1s}) $ can be expanded around the value $ \omega_f+E_{1s}=E_{np} $
\begin{eqnarray}
\label{27}
\widehat{\Sigma}(\omega_f+E_{1s})=\widehat{\Sigma}(E_{np})+(\omega_f+E_{1s}-E_{np})\widehat{\Sigma '}(E_{np})+...\;,
\end{eqnarray}
where $ \widehat{\Sigma '}(E_{np})\equiv\frac{d}{d\xi}\widehat{\Sigma}(\xi)|_{\xi=E_{np}} $. The first two terms of the expansion (\ref{27}) are ultraviolet divergent and require the renormalization. The methods of the renormolization in the bound electron QED are described, for example in \cite{45}. In order to obtain the line profile for the emission process we retain the first term of the expansion (\ref{27}) and consider the energy denominator in Eq. (\ref{25}) as a complex quantity:
\begin{eqnarray}
\label{28}
(\widehat{\Sigma}(E_{np}))_{np,np}=L^{SE}_{np}-\frac{i}{2}\Gamma_{np}\;.
\end{eqnarray}
Here $ L^{SE}_{np} $ is the electron self-energy contribution to the electron Lamb shift and $ \Gamma_{np} $ is the one-photon radiative level width \cite{44}. Apart from $ L^{SE} $ contribution
there is also the vacuum polarization $ L^{VP} $ contribution \cite{32}, but the vacuum polarization contribution is pure real and does not change the imaginary part in Eq. (\ref{28}). Now the emission amplitude reads
\begin{eqnarray}
\label{29}
U^{em}_{np-1s}=e\frac{(\gamma_{\mu}A^{*(\vec{k}_f\vec{e}_f)}_{\mu})_{1snp}}{\omega_f+E_{1s}-E_{np}-L_{np}+\frac{i}{2}\Gamma_{np}}\;,
\end{eqnarray}
where $ L_{np}=L^{SE}_{np}+L^{VP}_{np} $.

As a next step one has to take the amplitude Eq. (\ref{29}) by square modulus, then integrate over the photon emission directions $ \vec{\nu}_f $ and sum over the photon polarizations. We define the one-photon transition rate for the transition $ np-1s $ like
\begin{eqnarray}
\label{30}
W^{1\gamma}_{np-1s}=2\pi\omega^2_{res}\sum\limits_{\vec{e}_f}\int \frac{d\vec{\nu_f}}{(2\pi)^3}|(\gamma_{\mu}A^{*(\vec{k}_f,\vec{e}_f)}_{\mu})_{np1s}|^2\;,
\end{eqnarray}
where $ \omega_{res} $ is the resonant photon frequency. In Eq. (\ref{30}) it is assumed also the summation over the degenerate substates of the final state and averaging over the degenerate substates of the initial state. These operations we will not designate explicitly since it does not influence our argumentation. The same will concern the two-photon and three-photon transitions in the subsequent sections.

From Eq. (\ref{29}) we obtain for the absolute probability of the photon emission with the frequency in an interval between $ \omega_f $ and $ \omega_f+d\omega_f $
\begin{eqnarray}
\label{31}
dw_{np-1s}(\omega_f)=\frac{1}{2\pi}\frac{W^{1\gamma}_{np-1s}d\omega_f}{(\omega_f+E_{1s}-E_{np}-L_{np})^2+\frac{1}{4}\Gamma^2_{np}}\;.
\end{eqnarray}
Due to the factor $ \frac{1}{2\pi} $ the Lorentz profile Eq. (\ref{31}) is normalized to unity for the Lyman-alpha transition
\begin{eqnarray}
\label{32}
\int\limits_0^{\infty}dw_{2p-1s}=1\;.
\end{eqnarray}
In case $ n>2 $ the Lorentz profile is normalized to the branching ratio for the transition $ np-1s $:
\begin{eqnarray}
\label{33}
\int\limits_{-\infty}^{\infty}dw_{np-1s}=\frac{W^{1\gamma}_{np-1s}}{\Gamma_{np}}=b^{1\gamma}_{np-1s}\;.
\end{eqnarray}

The Lamb shift for the ground $ 1s $ state enters the energy denominator in Eq. (\ref{31}) in a different way. Insertions of the electron self-energy corrections in the outer electron lines in Fig. 1, unlike the insertions in an internal electron line lead to the singularities when the intermediate states in propagators are equal to $ 1s $. This singularities are not connected with the frequency resonances. To regularize these singularities one has to introduce Gel-Mann and Low \cite{46} adiabatic $ S $-matrix as it was done in \cite{47}. It was demonstrated that the summation of the infinite series of the singular in the adiabatic parameter $ \lambda $ terms can be converted to an exponential factor. The amplitude Eq. (\ref{25}) should be replaced by
\begin{eqnarray}
\label{34}
\lim\limits_{\lambda\rightarrow 0}U^{sc}_{1s(np)}(\lambda)=\lim\limits_{\lambda\rightarrow 0}e^2\frac{(\gamma_{\mu}A^{*(\vec{k}_f\vec{e}_f)}_{\mu})_{1snp}(\gamma_{\mu}A^{(\vec{k}_i\vec{e}_i)}_{\mu})_{np1s}}{\omega_f+E_{1s}+L_{1s}-E_{np}-L_{np}+\frac{i}{2}\Gamma_{np}}e^{-\frac{i}{\lambda}(\widehat{\Sigma}(E_{1s}))_{1s1s}}\;.
\end{eqnarray}
Since for the ground state the matrix element $ (\widehat{\Sigma}(E_{1s}))_{1s1s} $ is pure real, for the probability this gives
\begin{eqnarray}
\label{35}
\lim\limits_{\lambda\rightarrow 0}|e^{-\frac{i}{\lambda}(\widehat{\Sigma}(E_{1s}))_{1s1s}}|=1
\end{eqnarray}
and thus the Lamb shift $ L_{1s} $ arrives in the expression (\ref{31}) for the Lorentz profile.
Note, however, that if we apply Eq. (\ref{35}) to the excited state and take into account the width of the excited level we will obtain zero transition probability. Strictly speaking this means that it is incorrect to evaluate the transition probabilities via the nondiagonal $ S $-matrix elements as is usually done in QED for atoms and it is necessary to start with the process of excitation using the procedure described in the present paper. However in most cases Eq. (\ref{35}) can be ignored and the correct results for transitions rates are obtained, evaluating the square modules of the nondiagonal $ S $-matrix elements. Only in the special situations as in case of the multiphoton cascade transitions considered in the subsequent sections of the present paper, more refined analysis is required.

\section{Two-photon $ns-1s$ transition}

In this section we describe the two-photon transition to the ground state using as an example $ ns-1s $ two-photon transitions. According to our approach we have to start with the Feynman graph depicted in Fig. 2a. The two-photon resonant excitation is the most natural and convenient way to describe the excitation process in this case. The resonance condition is
\begin{eqnarray}
\label{36}
\omega_{i1}+\omega_{i2}=\omega_0^{ns}=E_{ns}-E_{1s}\;.
\end{eqnarray}

Constructing the $ S $-matrix element corresponding to the Feynman graph Fig. 2a, inserting the expressions for the electron and photon wave functions as well as the expressions for the electron propagators Eqs. (\ref{12})-(\ref{14}), integrating over time and frequency variables  and using Eq. (\ref{16}) for the scattering amplitude results
\begin{eqnarray}
\label{37}
U^{(4)sc.}_{1s}=e^4\sum\limits_{n_1n_2n_3}\frac{(\gamma_{\mu_1}A_{\mu_1}^{*(\vec{k}_{f_2}\vec{e}_{f_2})})_{1sn_1}(\gamma_{\mu_2}A_{\mu_2}^{*(\vec{k}_{f_1}\vec{e}_{f_1})})_{n_1n_2}}{(\omega_{f_2}+E_{1s}-E_{n_1})(\omega_{f_2}+\omega_{f_1}+E_{1s}-E_{n_2})}\times\\\nonumber
\frac{(\gamma_{\mu_3}A_{\mu_3}^{(\vec{k}_{i_2}\vec{e}_{i_2})})_{n_2n_3}(\gamma_{\mu_4}A_{\mu_4}^{(\vec{k}_{i_1}\vec{e}_{i_1})})_{n_31s}}{(\omega_{f_2}+\omega_{f_1}-\omega_{i_2}+E_{1s}-E_{n_3})}\;.
\end{eqnarray}
The energy conservation in this process is implemented by the condition
\begin{eqnarray}
\label{38}
\omega_{f_1}+\omega_{f_2}=\omega_{i_1}+\omega_{i_2}
\end{eqnarray}
and the resonance condition is given by Eq. (\ref{36}). From Eq. (\ref{36}) follows the approximate energy conservation law similar to Eq. (\ref{20b})
\begin{eqnarray}
\label{39a}
|\omega_{f_1}+\omega_{f_2}-E_{ns}+E_{1s}|\leqslant \Gamma_{ns}\;,
\end{eqnarray}\
which can be replaced by equation similar to Eq. (\ref{20c})
\begin{eqnarray}
\label{39b}
\omega_{f_1}+\omega_{f_2}=E_{ns}-E_{1s}\;,
\end{eqnarray}
when $ \Gamma_{ns} $ can be neglected.
According to Eqs. (\ref{36}) and (\ref{38}) the last energy denominator in Eq. (\ref{37}) can be replaced by 
\begin{eqnarray}
\label{39}
\omega_{f_2}+\omega_{f_1}-\omega_{i_2}+E_{1s}-E_{n_3}=\omega_{i_1}+E_{1s}-E_{n_3}\;,
\end{eqnarray}
i.e. does not depend on the frequencies of emitted photons.

In the resonance approximation we retain only one term $ n_2=ns $ in the sum over $ n_2 $ which yields

\begin{eqnarray}
\label{40}
U^{(4)sc}_{1s(ns)}=e^4\sum\limits_{n_1}\frac{(\gamma_{\mu_1}A_{\mu_1}^{*(\vec{k}_{f_2}\vec{e}_{f_2})})_{1sn_1}(\gamma_{\mu_2}A_{\mu_2}^{*(\vec{k}_{f_1}\vec{e}_{f_1})})_{n_1ns}}{(\omega_{f_2}+E_{1s}-E_{n_1})(\omega_{f_2}+\omega_{f_1}+E_{1s}-E_{ns})}\sum\limits_{n_3}\frac{(\gamma_{\mu_3}A_{\mu_3}^{(\vec{k}_{i_2}\vec{e}_{i_2})})_{nsn_3}(\gamma_{\mu_4}A_{\mu_4}^{(\vec{k}_{i_1}\vec{e}_{i_1})})_{n_31s}}{(\omega_{i_1}+E_{1s}-E_{n_3})}\;.
\end{eqnarray}
Starting from Eq. (\ref{40}) we can write down the expression for the two-photon emission amplitude as
\begin{eqnarray}
\label{41}
U^{(2)em}_{ns-1s}=e^2\sum\limits_{n_1}\frac{(\gamma_{\mu_1}A_{\mu_1}^{*(\vec{k}_{f_2}\vec{e}_{f_2})})_{1sn1}(\gamma_{\mu_2}A_{\mu_2}^{*(\vec{k}_{f_1}\vec{e}_{f_1})})_{n_1ns}}{(\omega_{f_2}+E_{1s}-E_{n_1})(\omega_{f_2}+\omega_{f_1}+E_{1s}-E_{ns})}
\end{eqnarray}
with the condition Eq. (\ref{36}) remaining valid for the frequencies $\omega_{f_1}, \omega_{f_2}$ due to Eq. (\ref{38}).

Up to now all formulas above in this section were valid for any $ ns $ levels, beginning from $ n=2s $. Now we have to take into account the form of the resonance produced by the second energy denominator in Eq. (\ref{41}). The width of this resonance for $ 2s $ level is defined by the two-photon transition $ 2s\rightarrow 1s+2\gamma $. This width should arrive as the imaginary part of the matrix element of the second-order electron self-energy operator, i.e. from two-loop insertions to the Feynman graph Fig. 2a. 

For $ n>2 $ there is always leading one-photon contribution to the total width $ \Gamma_{ns} $, for example the $ W^{1\gamma}_{3s-2p} $ transition rate in case $ n=3 $. Assuming the existence of such a contribution we will continue our studies by inserting the one-loop electron self-energy corrections to the central propagator in Fig. 1a (the Low procedure). The first term of the Low sequence is depicted in Fig. 2b.

Returning back to the scattering amplitude Eq. (\ref{39}) and proceeding along the same way as in the case of the one-photon decay we obtain an expression similar to Eq. (\ref{23}) in the one-photon case:
\begin{eqnarray}
\label{42}
U^{(6)sc}_{1s(ns)}=e^4\sum\limits_{n_1n_2n_3}\frac{(\gamma_{\mu_1}A_{\mu_1}^{*(\vec{k}_{f_2}\vec{e}_{f_2})})_{1sn_1}(\gamma_{\mu_2}A_{\mu_2}^{*(\vec{k}_{f_1}\vec{e}_{f_1})})_{n_1n_2}}{(\omega_2+E_{1s}-E_{n_1})(\omega_{f_2}+\omega_{f_1}+E_{1s}-E_{n_2})}\times\\\nonumber
\frac{(\widehat{\Sigma}(\omega_{f_1}+\omega_{f_2}+E_{1s}))_{n_2n_3}}{(\omega_{f_2}+\omega_{f_1}+E_{1s}-E_{n_3})}\times\frac{(\gamma_{\mu_3}A_{\mu_3}^{(\vec{k}_{i_2}\vec{e}_{i_2})})_{n_3n_3}(\gamma_{\mu_4}A_{\mu_4}^{(\vec{k}_{i_1}\vec{e}_{i_1})})_{n_41s}}{(\omega_1+E_{1s}-E_{n_4})}\;.
\end{eqnarray}
In the resonance approximation setting $ n_2=n_3=ns $ we have
\begin{eqnarray}
\label{43}
U^{(6)sc}_{1s(ns)}=U^{(4)sc}_{1s(ns)}\frac{(\widehat{\Sigma}(\omega_{f_1}+\omega_{f_2}+E_{1s}))_{nsns}}{(\omega_{f_2}+\omega_{f_1}+E_{1s}-E_{ns})}\;.
\end{eqnarray}
Producing further the Low sequence and performing the summation of the arising geometric progression results
\begin{eqnarray}
\label{44}
U^{sc}_{1s(ns)}=e^4\sum\limits_{n_1}\frac{(\gamma_{\mu_1}A_{\mu_1}^{*(\vec{k}_{f_2}\vec{e}_{f_2})})_{n_1ns}(\gamma_{\mu_2}A_{\mu_2}^{*(\vec{k}_{f_1}\vec{e}_{f_1})})_{n_1n_2}}{(\omega_{f_2}+E_{1s}-E_{n_1})}\times\\\nonumber
\frac{1}{\omega_{f_2}+\omega_{f_1}+E_{1s}-E_{ns}-(\widehat{\Sigma}(\omega_{f_1}+\omega_{f_2}+E_{1s}))_{ns,ns}}\sum\limits_{n_2}\frac{(\gamma_{\mu_3}A_{\mu_3}^{(\vec{k}_{i_2}\vec{e}_{i_2})})_{n_1ns}(\gamma_{\mu_4}A_{\mu_4}^{(\vec{k}_{i_1}\vec{e}_{i_1})})_{n_1n_2}}{(\omega_{i_1}+E_{1s}-E_{n_2})}\;.
\end{eqnarray}
The emission amplitude for the two-photon decay process $ ns\rightarrow 1s+2\gamma $ looks like
\begin{eqnarray}
\label{45}
U^{em}_{ns-1s}=e^2\sum\limits_{n_1}\frac{(\gamma_{\mu_1}A_{\mu_1}^{*(\vec{k}_{f_2}\vec{e}_{f_2})})_{n_1ns}(\gamma_{\mu_2}A_{\mu_2}^{*(\vec{k}_{f_1}\vec{e}_{f_1})})_{n_1n_2}}{(\omega_{f_2}+E_{1s}-E_{n_1})}\times\\\nonumber
\frac{1}{\omega_{f_2}+\omega_{f_1}+E_{1s}-E_{ns}-(\widehat{\Sigma}(\omega_{f_1}+\omega_{f_2}+E_{1s}))_{ns,ns}}\;.
\end{eqnarray}
At the point of the resonance we expand the operator 
\begin{eqnarray}
\label{46}
\widehat{\Sigma}(\omega_{f_1}+\omega_{f_2}+E_{1s})=\widehat{\Sigma}(E_{ns})+...
\end{eqnarray}
and using the equality
\begin{eqnarray}
\label{47}
(\widehat{\Sigma}(E_{ns}))_{ns,ns}=L^{SE}_{ns}-\frac{i}{2}\Gamma_{ns}\;,
\end{eqnarray}
arrive at
\begin{eqnarray}
\label{48}
U^{em}_{ns-1s}=e^2\sum\limits_{n_1}\frac{(\gamma_{\mu_1}A_{\mu_1}^{*(\vec{k}_{f_2}\vec{e}_{f_2})})_{n_1ns}(\gamma_{\mu_2}A_{\mu_2}^{*(\vec{k}_{f_1}\vec{e}_{f_1})})_{n_1n_2}}{(\omega_{f_2}+E_{1s}-E_{n_1})(\omega_{f_2}+\omega_{f_1}+E_{1s}-E_{ns}+\frac{i}{2}\Gamma_{ns})}\;.
\end{eqnarray}

In Eq. (\ref{48}) we have omitted the Lamb shift of the $ ns $ level in the second energy denominator. In what follows the Lamb shift will play no significant role in our derivations.

The value $ \Gamma_{ns} $ is defined in a different way for the different $ ns $ states. For example,  $\Gamma_{3s}=W^{1\gamma}_{3s-2p}$ since there are no other one-photon decay channels for $ 3s $ level. The further investigations of the two-photon transition probabilities should be performed separately for different $ n $. In the next section we will continue these investigations for $ 3s\rightarrow1s+2\gamma $ transition.
\section{Two-photon $3s-1s$ transition}

The further studies of the $ 3s-1s $ transition we can start with the expression for the emission amplitude Eq. (\ref{48}) written for the case $ ns=3s $. The Feynman graphs for the resonance two-photon scattering with the excitation of $ 3s $ level are depicted in Fig. 3. To the expression Eq. (\ref{48}) we have to add also another term corresponding to the Feynman graph Fig. 3a with the interchanged positions of the $ \vec{k}_{f_1},\vec{e}_{f_1} $ and $ \vec{k}_{f_2},\vec{e}_{f_2} $ photons. This yields
\begin{eqnarray}
\label{49}
U^{em}_{3s-1s}=e^2\sum\limits_{n_1}\left\lbrace \frac{(\gamma_{\mu_1}A_{\mu_1}^{*(\vec{k}_{f_2}\vec{e}_{f_2})})_{1sn_1}(\gamma_{\mu_2}A_{\mu_2}^{*(\vec{k}_{f_1}\vec{e}_{f_1})})_{n_13s}}{\omega_{f_2}+E_{1s}-E_{n_1}}+\frac{(\gamma_{\mu_1}A_{\mu_1}^{*(\vec{k}_{f_1}\vec{e}_{f_1})})_{1sn_1}(\gamma_{\mu_2}A_{\mu_2}^{*(\vec{k}_{f_2}\vec{e}_{f_2})})_{n_13s}}{\omega_{f_1}+E_{1s}-E_{n_1}}\right\rbrace\times\\\nonumber\times \frac{1}{\omega_{f_2}+\omega_{f_1}+E_{1s}-E_{3s}+\frac{i}{2}\Gamma_{3s}}\;.
\end{eqnarray}

For the $ 3s-1s $ two-photon transition only one cascade is possible: $ 3s-2p-1s $. Accordingly, the two new resonance conditions arise (these resonances were defined in section I):
\begin{eqnarray}
\label{50}
\omega^{res.1}=E_{3s}-E_{2p}\;,
\end{eqnarray}
\begin{eqnarray}
\label{51}
\omega^{res.2}=E_{2p}-E_{1s}\;.
\end{eqnarray}

Consider first cascade contribution to Eq. (\ref{49}). For this purpose we have to set $ n_1=2p $. Then
\begin{eqnarray}
\label{52}
U^{em,\; cascade}_{3s-2p-1s}=e^2\left\lbrace \frac{(\gamma_{\mu_1}A_{\mu_1}^{*(\vec{k}_{f_2}\vec{e}_{f_2})})_{1s2p}(\gamma_{\mu_2}A_{\mu_2}^{*(\vec{k}_{f_1}\vec{e}_{f_1})})_{2p3s}}{\omega_{f_2}+E_{1s}-E_{2p}}+\frac{(\gamma_{\mu_1}A_{\mu_1}^{*(\vec{k}_{f_1}\vec{e}_{f_1})})_{1s2p}(\gamma_{\mu_2}A_{\mu_2}^{*(\vec{k}_{f_2}\vec{e}_{f_2})})_{2p3s}}{\omega_{f_1}+E_{1s}-E_{2p}}\right\rbrace\times\\\nonumber\times \frac{1}{\omega_{f_2}+\omega_{f_1}+E_{1s}-E_{3s}+\frac{i}{2}\Gamma_{3s}}\;.
\end{eqnarray}

The first term in the curly  brackets describes the resonance (\ref{50}), the second term describes the resonance (\ref{51}) (see Appendix A). Applying the Low procedure (insertions and summation of the infinite chain of the electron self-energy corrections) to the upper electron propagators in Fig. 3b we find
\begin{eqnarray}
\label{53}
U^{em,\; cascade}_{3s-2p-1s}=e^2\left\lbrace \frac{(\gamma_{\mu_1}A_{\mu_1}^{*(\vec{k}_{f_2}\vec{e}_{f_2})})_{1s2p}(\gamma_{\mu_2}A_{\mu_2}^{*(\vec{k}_{f_1}\vec{e}_{f_1})})_{2p3s}}{\omega_{f_2}+E_{1s}-E_{2p}+\frac{i}{2}\Gamma_{2p}}+\frac{(\gamma_{\mu_1}A_{\mu_1}^{*(\vec{k}_{f_1}\vec{e}_{f_1})})_{1s2p}(\gamma_{\mu_2}A_{\mu_2}^{*(\vec{k}_{f_2}\vec{e}_{f_2})})_{2p3s}}{\omega_{f_1}+E_{1s}-E_{2p}+\frac{i}{2}\Gamma_{2p}}\right\rbrace \times\\\nonumber\times\frac{1}{\omega_{f_2}+\omega_{f_1}+E_{1s}-E_{3s}+\frac{i}{2}\Gamma_{3s}}\;.
\end{eqnarray}

Now we take $ U^{em, cascade}_{3s-2p-1s} $ by square modulus, integrate over the emitted photons directions and sum over the polarizations of both photons. The formula (\ref{30}) should be used for presentation of the results of these integrations and summation via the one-photon transition rates. Consider first the square modulus of the first term in the curly brackets and the factor outside the curly brackets in Eq. (\ref{53}). This term is represented by Fig. 3a and corresponds to the contribution of the resonance 1 in Eq. (\ref{50}). In this case we are interested to derive the Lorentz line profile for the upper link of the cascade $ 3s-2p-1s $. Therefore we have to integrate first over frequency of the second emitted photon, i.e. $ \omega_{f_2} $. In principle the integration over both photon frequencies should be done with Eq. (\ref{39a}) taken into account, i.e.
\begin{eqnarray}
\label{54a}
\int\limits_0^{\omega_{max}}d\omega_{f_1}\int\limits_0^{\omega_{1}}d\omega_{f_2}=\frac{1}{2}\int\limits_0^{\omega_{max}}d\omega_{f_1}\int\limits_0^{\omega_{max}}d\omega_{f_2}\;,
\end{eqnarray}
where $ \omega_{max}=E_{2s}-E_{1s} $.

Eq. (\ref{54a}) holds due to the symmetry of Eq. (\ref{53}) with respect to permutation $ \omega_{f_1}\leftrightarrows \omega_{f_2} $.

Here, the integration over the frequency $ \omega_{f_2} $ in Eq. (\ref{53}) we perform in the complex plane. Since only the pole terms contribute we can extend the interval of integration to $ (-\infty, +\infty) $ and not to refer to Eq. (\ref{39a}) or (\ref{54a}). Then using Cauchy theorem after some algebraic transformation (see for details the Appendix A) we obtain the cascade contribution (resonance 1) to the differential branching ratio
\begin{eqnarray}
\label{54}
db^{2\gamma(resonance\;1)}_{3s-2p-1s}(\omega)=\frac{1}{2\pi}\frac{\Gamma_{3s}+\Gamma_{2p}}{\Gamma_{3s}\Gamma_{2p}}\frac{W_{3s-2p}^{1\gamma}(\omega^{res. 1})W_{2p-1s}^{1\gamma}(\omega^{res. 2})d\omega}{(\omega-\omega^{res. 1})^2+\frac{1}{4}(\Gamma_{3s}+\Gamma_{2p})^2}\;
\end{eqnarray}
(here we have changed the notation for the frequency from  $ \omega_{f_1} $ to $ \omega $).

The differential branching ratio $ db^{2\gamma} $ is connected with the differential transition rate $ dw^{2\gamma}_{ns-1s}(\omega) $ via
\begin{eqnarray}
\label{55}
db^{2\gamma}_{ns-1s}(\omega)=\frac{dw^{2\gamma}_{ns-1s}}{\Gamma_{ns}}\;.
\end{eqnarray}

This definition concerns not only the cascade contributions but all the contributions in Eq. (\ref{2}) for the two-photon decay of any $ ns $-state:
\begin{eqnarray}
\label{56}
db^{2\gamma}_{ns-1s}=db^{2\gamma(cascade)}_{ns-1s}+db^{2\gamma(pure)}_{ns-1s}+db^{2\gamma(interference)}_{ns-1s}=\frac{1}{\Gamma_{ns}}(dw^{2\gamma(cascade)}_{ns-1s}+dw^{2\gamma(pure)}_{ns-1s}+dw^{2\gamma(interference)}_{ns-1s})\;.
\end{eqnarray}

Combining now the formulas (\ref{54}), (\ref{55}) we arrive at the expression (\ref{4}) presented in the Introduction. The integration of Eq. (\ref{56}) over the remaining frequency will give the total branching ratio
\begin{eqnarray}
\label{57}
b^{2\gamma}_{ns-1s}=\frac{W^{2\gamma}_{ns-1s}}{\Gamma_{ns}}\;.
\end{eqnarray}

Note that this last integration according to Eq. (\ref{54a}) should be done within the interval ($ 0 $, $ \omega_{max} $) since now no pole approximation can be used.

The second term in the curly brackets in Eq. (\ref{53}) is represented by the Feynman graph Fig. 3a (with the change of the photons $ \omega_{f_1}\leftrightarrows\omega_{f_2} $) and corresponds to the resonance 2 in Eq. (\ref{51}), i.e. to the lower link of cascade. To obtain the Lorentz profile for this lower link we have to integrate over the frequency of the first emitted photon, i.e. again over $ \omega_{f_2} $ after taking the square modulus of this term and the factor outside the curly brackets in Eq. (\ref{53}). Replacing notation $ \omega_{f_1} $ to $ \omega  $ we obtain the cascade contribution (resonance 2) to the differential branching ratio:
\begin{eqnarray}
\label{58}
db^{2\gamma(resonance 2)}_{3s-2p-1s}(\omega)=\frac{1}{2\pi}\frac{1}{\Gamma_{3s}}\frac{W_{3s-2p}^{1\gamma}(\omega^{res. 1})W_{2p-1s}^{1\gamma}(\omega^{res. 2})d\omega}{(\omega-\omega^{res. 2})^2+\frac{1}{4}\Gamma_{2p}^2}\;.
\end{eqnarray}
Combining the formulas (\ref{55}) and (\ref{58}) we arrive at the expression (\ref{5}) presented in the Introduction.

The interference between two terms in Eq. (\ref{53}) should not be taken into account since these two terms correspond to the resonances located far from each other: at the distance $ \omega_{max} $ in the frequency scale.
\section{"Pure two-photon" and interference contributions}

Returning to Eq. (\ref{49}) we consider this expression with the state $ 2p $ excluded from the summation over $ n_1 $ as a "pure two-photon" contribution to the transition amplitude $ 3s-1s $. The words "pure two-photon"\;should not be understood literally: we have to remember that the exact separation of this "pure two-photon"\;contribution is not possible \cite{34}. This exclusion corresponds to the "pole approximation" employed in Section IV for the description of the cascade contribution: extension of the of the first frequency integration over the interval ($ -\infty,\infty $). In \cite{34} the more general approach was developed, when the resonances were regularized only within the "windows" of the different breadth. Then the $ 2p $ state should be eliminated from the sum over $ n_1 $ only within "windows". The "pole approximation" corresponds to the window breadth $[\omega]=\infty$.

Since the energy denominators (apart from the factor outside the curly brackets in (\ref{49})) now become nonsingular we can employ the energy conservation law Eq. (\ref{38}) to replace the frequency $ \omega_{f_1} $ in the second denominator in curly brackets in Eq. (\ref{49}) by $ \omega_{f_1}=\omega_{max}-\omega_{f_2} $. Then the "pure two-photon" contribution to the amplitude becomes

\begin{eqnarray}
\label{59}
U^{em., pure}_{3s-1s}=e^2\sum\limits_{n_1\neq 2p}\left\lbrace \frac{(\gamma_{\mu_1}A_{\mu_1}^{*(\vec{k}_{f_2}\vec{e}_{f_2})})_{1sn_1}(\gamma_{\mu_2}A_{\mu_2}^{*(\vec{k}_{f_1}\vec{e}_{f_1})})_{n_13s}}{\omega_{f_2}+E_{1s}-E_{n_1}}+\frac{(\gamma_{\mu_1}A_{\mu_1}^{*(\vec{k}_{f_1}\vec{e}_{f_1})})_{1sn_1}(\gamma_{\mu_2}A_{\mu_2}^{*(\vec{k}_{f_2}\vec{e}_{f_2})})_{n_13s}}{E_{3s}-\omega_{f_2}-E_{n_1}}\right\rbrace\times\\\nonumber \frac{1}{\omega_{f_2}+\omega_{f_1}+E_{1s}-E_{3s}+\frac{i}{2}\Gamma_{3s}}\;.
\end{eqnarray}
Taking Eq. (\ref{59}) by square modulus, integrating over the directions of the emitted photons, summing over the polarizations, integrating over $ \omega_{f_1} $, and changing the notation $ \omega_{f_2}=\omega $ results
\begin{eqnarray}
\label{60}
db^{2\gamma(pure)}_{3s-1s}(\omega)=e^4\omega^2(\omega_0-\omega)^2\sum\limits_{\vec{e}_{f_1}}\sum\limits_{\vec{e}_{f_2}}\int \frac{d\vec{\nu_{f_1}}}{(2\pi)^3}\frac{d\vec{\nu_{f_2}}}{(2\pi)^3}\times\nonumber\\\left|\sum\limits_{n_1\neq 2p}\left\lbrace \frac{(\gamma_{\mu_1}A_{\mu_1}^{*(\vec{k}_{f_2}\vec{e}_{f_2})})_{1sn_1}(\gamma_{\mu_2}A_{\mu_2}^{*(\vec{k}_{f_1}\vec{e}_{f_1})})_{n_13s}}{\omega+E_{1s}-E_{n_1}}+\frac{(\gamma_{\mu_1}A_{\mu_1}^{*(\vec{k}_{f_1}\vec{e}_{f_1})})_{1sn_1}(\gamma_{\mu_2}A_{\mu_2}^{*(\vec{k}_{f_2}\vec{e}_{f_2})})_{n_13s}}{E_{3s}-\omega-E_{n_1}}\right\rbrace\;\right|^2 \frac{1}{\Gamma_{3s}}d\omega\;.
\end{eqnarray}
Then, according to Eq. (\ref{56}) the "pure two-photon" contribution to the differential transition rate is
\begin{eqnarray}
\label{61}
dW^{2\gamma(pure)}_{3s-1s}(\omega)=e^4\omega^2(\omega_0-\omega)^2\sum\limits_{\vec{e}_1}\sum\limits_{\vec{e}_2}\int \frac{d\vec{\nu_{f_1}}}{(2\pi)^3}\frac{d\vec{\nu_{f_2}}}{(2\pi)^3}\times\nonumber\\\left|\sum\limits_{n_1\neq 2p}\left\lbrace \frac{(\gamma_{\mu_1}A_{\mu_1}^{*(\vec{k}_{f_2}\vec{e}_{f_2})})_{1sn_1}(\gamma_{\mu_2}A_{\mu_2}^{*(\vec{k}_{f_1}\vec{e}_{f_1})})_{n_13s}}{\omega+E_{1s}-E_{n_1}}+\frac{(\gamma_{\mu_1}A_{\mu_1}^{*(\vec{k}_{f_1}\vec{e}_{f_1})})_{1sn_1}(\gamma_{\mu_2}A_{\mu_2}^{*(\vec{k}_{f_2}\vec{e}_{f_2})})_{n_13s}}{E_{3s}-\omega-E_{n_1}}\right\rbrace\;\right|^2 d\omega\;.
\end{eqnarray}
Now, using Eq. (\ref{53}) and Eq. (\ref{59}) we can write down the interference contribution to the differential branching ratio as
\begin{eqnarray}
\label{62} 
db^{2\gamma(interference)}_{3s-1s}=2Re\sum\limits_{\vec{e}_1}\sum\limits_{\vec{e}_2}\int \frac{d\vec{\nu_{f_1}}}{(2\pi)^3}\int\frac{d\vec{\nu_{f_2}}}{(2\pi)^3}\int d\omega_{f_1}\omega_{f_1}^2\omega_{f_2}^2U^{em(pure)*}_{3s-1s}U^{em.(cascade)}=\\\nonumber =2Re\;e^4 \sum\limits_{\vec{e}_1}\sum\limits_{\vec{e}_2}\int\frac{d\vec{\nu_{f_1}}}{(2\pi)^3}\int \frac{d\vec{\nu_{f_2}}}{(2\pi)^3}\int d\omega_{f_1}\omega_{f_1}^2\omega_{f_2}^2\times\\\nonumber\left( \sum\limits_{n_1\neq 2p}\left\lbrace\frac{(\gamma_{\mu_1}A_{\mu_1}^{*(\vec{k}_{f_2}\vec{e}_{f_2})})_{1sn_1}(\gamma_{\mu_2}A_{\mu_2}^{*(\vec{k}_{f_1}\vec{e}_{f_1})})_{n_13s}}{\omega_{f_2}+E_{1s}-E_{n_1}}+\frac{(\gamma_{\mu_1}A_{\mu_1}^{*(\vec{k}_{f_1}\vec{e}_{f_1})})_{1sn_1}(\gamma_{\mu_2}A_{\mu_2}^{*(\vec{k}_{f_2}\vec{e}_{f_2})})_{n_13s}}{E_{3s}-E_{n_1}-\omega_{f_2}}\right\rbrace\right)^*\times\\\nonumber\left\lbrace \frac{(\gamma_{\mu_1}A_{\mu_1}^{*(\vec{k}_{f_2}\vec{e}_{f_2})})_{1sn_1}(\gamma_{\mu_2}A_{\mu_2}^{*(\vec{k}_{f_1}\vec{e}_{f_1})})_{n_13s}}{\omega_{f_2}+E_{1s}-E_{2p}+\frac{i}{2}\Gamma_{2p}}+\frac{(\gamma_{\mu_1}A_{\mu_1}^{*(\vec{k}_{f_1}\vec{e}_{f_1})})_{1sn_1}(\gamma_{\mu_2}A_{\mu_2}^{*(\vec{k}_{f_2}\vec{e}_{f_2})})_{n_13s}}{\omega_{f_1}+E_{1s}-E_{2p}+\frac{i}{2}\Gamma_{2p}}\right\rbrace\times\\\nonumber\frac{d\omega_{f_2}}{(\omega_{f_1}+\omega_{f_2}+E_{1s}-E_{3s})^2+\frac{1}{4}\Gamma_{3s}^2}\;.
\end{eqnarray}

The integration in the complex $ \omega_{f_1} $ plane can be extended over the entire interval $ -\infty\leqslant\omega_{f_1}\leqslant+\infty $ since only the pole term contributes; then we have to take the real part of the expression obtained.

The "pure two-photon" amplitude in Eq. (\ref{62}) we can assume to be pure real.

Now, using Eq. (\ref{55}) for the interference contribution to the differential two-photon transition rate $ 3s-1s $ we find (changing notation $ \omega_{f_2} $ to $ \omega $)

\begin{eqnarray}
\label{63}
dW^{2\gamma(interference)}_{3s-1s}(\omega)=\frac{2(\omega-\omega^{res2})F_1^{3s}(\omega)}{(\omega-\omega^{res2})^2+\frac{1}{4}(\Gamma_{3s}+\Gamma_{2p})^2}d\omega+\frac{2(\omega-\omega^{res1})F_2^{3s}(\omega)}{(\omega-\omega^{res1})^2+\frac{1}{4}\Gamma_{2p}^2}d\omega\;.
\end{eqnarray}

\begin{eqnarray}
\label{64}
F^{3s}_i(\omega)=\sum\limits_{\vec{e}_{f_1}}\sum\limits_{\vec{e}_{f_2}}\int \frac{d\vec{\nu_{f_1}}}{(2\pi)^3}\int\frac{d\vec{\nu_{f_2}}}{(2\pi)^3}f^{3s}(\omega)\varphi_{i}^{3s}, i=1,2\;.
\end{eqnarray}

\begin{eqnarray}
\label{65}
f^{3s}=\omega^2(\omega^{3s}_0-\omega)^2\sum\limits_{n_1\neq 2p}\left\lbrace \frac{(\gamma_{\mu_1}A_{\mu_1}^{*(\vec{k}_{f_2}\vec{e}_{f_2})})_{1sn_1}(\gamma_{\mu_2}A_{\mu_2}^{*(\vec{k}_{f_1}\vec{e}_{f_1})})_{n_13s}}{(\omega+E_{1s}-E_{n_1})}+\frac{(\gamma_{\mu_1}A_{\mu_1}^{*(\vec{k}_{f_1}\vec{e}_{f_1})})_{1sn_1}(\gamma_{\mu_2}A_{\mu_2}^{*(\vec{k}_{f_2}\vec{e}_{f_2})})_{n_13s}}{(E_{3s}-E_{n_1}-\omega)}\right\rbrace\;,
\end{eqnarray}
\begin{eqnarray}
\label{66}
\varphi_{1}^{3s}=(\gamma_{\mu_1}A_{\mu_1}^{*(\vec{k}_{f_2}\vec{e}_{f_2})})_{1s2p}(\gamma_{\mu_2}A_{\mu_2}^{*(\vec{k}_{f_1}\vec{e}_{f_1})})_{2p3s}\;,
\end{eqnarray}

\begin{eqnarray}
\label{67}
\varphi_{2}^{3s}=(\gamma_{\mu_1}A_{\mu_1}^{*(\vec{k}_{f_1}\vec{e}_{f_1})})_{1s2p}(\gamma_{\mu_2}A_{\mu_2}^{*(\vec{k}_{f_2}\vec{e}_{f_2})})_{2p3s}\;.
\end{eqnarray}
In Eqs. (\ref{65}) - (\ref{67}) it is assumed that $ |k_{f_2}|\equiv\omega $, $ |k_{f_1}|\equiv(\omega^{3s}_0-\omega) $.

\section{Two-photon $4s-1s$ transition}
Repeating the derivations for the $ 3s-1s $ transition for the case of $ 4s-1s $ transition we present first the contributions for the two cascades: $ 4s-2p-1s $ and $ 4s-3p-1s $:

1) Contribution from the upper link $ 4s-2p $ of the cascade $ 4s-2p-1s $:
\begin{eqnarray}
\label{68}
dW^{2\gamma(resonance 1)}_{4s-2p-1s}=\frac{1}{2\pi}\frac{\Gamma_{4s}+\Gamma_{2p}}{\Gamma_{2p}}\frac{W_{4s-2p}^{1\gamma}(\omega^{res. 1})W_{2p-1s}^{1\gamma}(\omega^{res. 2})d\omega}{(\omega-\omega^{res. 1})^2+\frac{1}{4}(\Gamma_{4s}+\Gamma_{2p})^2}\;,
\end{eqnarray}
where
\begin{eqnarray}
\label{69}
\omega^{res. 1}=E_{4s}-E_{2p},\;\; \omega^{res. 2}=E_{2p}-E_{1s}\;.
\end{eqnarray}
2) Contribution of the lower link $ 2p-1s $ of the cascade $ 4s-2p-1s $:
\begin{eqnarray}
\label{70}
dW^{2\gamma(resonance 1)}_{4s-2p-1s}=\frac{W_{4s-2p}^{1\gamma}(\omega^{res. 1})W_{2p-1s}^{1\gamma}(\omega^{res. 2})d\omega}{(\omega-\omega^{res. 1})^2+\frac{1}{4}\Gamma_{2p}^2}\;.
\end{eqnarray}
3) Contribution from the upper link $ 4s-3p $ of the cascade $ 4s-3p-1s $:
\begin{eqnarray}
\label{71}
dW^{2\gamma(resonance 1)}_{4s-3p-1s}=\frac{1}{2\pi}\frac{\Gamma_{4s}+\Gamma_{3p}}{\Gamma_{3p}}\frac{W_{4s-3p}^{1\gamma}(\omega^{res. 3})W_{3p-1s}^{1\gamma}(\omega^{res. 4})d\omega}{(\omega-\omega^{res. 3})^2+\frac{1}{4}(\Gamma_{4s}+\Gamma_{3p})^2}\;,
\end{eqnarray}
where
\begin{eqnarray}
\label{72}
\omega^{res. 3}=E_{4s}-E_{3p},\;\; \omega^{res. 4}=E_{3p}-E_{1s}\;.
\end{eqnarray}
4) Contribution of the lower link $ 3p-1s $ of the cascade $ 4s-3p-1s $:
\begin{eqnarray}
\label{73}
dW^{2\gamma(resonance 1)}_{4s-3p-1s}=\frac{W_{4s-3p}^{1\gamma}(\omega^{res. 3})W_{3p-1s}^{1\gamma}(\omega^{res. 4})d\omega}{(\omega-\omega^{res. 4})^2+\frac{1}{4}\Gamma_{3p}^2}\;.
\end{eqnarray}
Insertion of Eqs. (\ref{68}), (\ref{70}), (\ref{71}), (\ref{73}) in Eq. (\ref{1}) results:
\begin{eqnarray}
\label{74}
W^{2\gamma(cascade)}_{4s-1s}=\frac{1}{2}\int\limits_0^{\omega_0}\sum\limits_{i=1}^4dW^{2\gamma(resonance\; i)}=W^{1\gamma}_{4s-2p}+\frac{W^{1\gamma}_{3p-1s}}{\Gamma_{3p}}W^{1\gamma}_{4s-3p}=W^{1\gamma}_{4s-2p}+b^{1\gamma}_{3p-1s}W^{1\gamma}_{4s-3p}\;,
\end{eqnarray}
where $\;\omega_0=E_{4s}-E_{1s} $, $ b^{1\gamma}_{3p-1s} $ is the branching ratio for the transition $ 3p-1s $. We took into account that $ \Gamma_{3p}=W^{1\gamma}_{3p-1s}+W^{1\gamma}_{3p-2s} $ and $ b^{1\gamma}_{3p-1s}=\frac{W^{1\gamma}_{3p-1s}}{W^{1\gamma}_{3p-1s}+W^{1\gamma}_{3p-2s}} $.

Hence 
\begin{eqnarray}
\label{75}
W^{2\gamma(cascade)}_{4s-1s}\neq\Gamma_{4s}\;,
\end{eqnarray}
where
\begin{eqnarray}
\label{76}
\Gamma_{4s} =W^{1\gamma}_{4s-2p}+W^{1\gamma}_{4s-3p}\;,
\end{eqnarray}
unlike Eq. (\ref{7a}) in case of $ 3s-1s $ transition.
The "pure two-photon" contribution to the $ 4s-1s $ two-photon differential decay rate looks similar to the $ 3s-1s $ case (see Eq. (\ref{61})):
\begin{eqnarray}
\label{77}
dW^{2\gamma(pure)}_{4s-1s}(\omega)=e^4\omega^2(\omega_0-\omega)^2\sum\limits_{\vec{e}_{f_1}}\sum\limits_{\vec{e}_{f_2}}\int \frac{d\vec{\nu}_{f_1}}{(2\pi)^3}\frac{d\vec{\nu}_{f_2}}{(2\pi)^3}\times\\\nonumber\left|\sum\limits_{n_1\neq 2p,3p}\left\lbrace \frac{(\gamma_{\mu_1}A_{\mu_1}^{*(\vec{k}_{f_2}\vec{e}_{f_2})})_{1sn_1}(\gamma_{\mu_2}A_{\mu_2}^{*(\vec{k_{f_1}}\vec{e_{f_1}})})_{n_14s}}{\omega+E_{1s}-E_{n_1}}+\frac{(\gamma_{\mu_1}A_{\mu_1}^{*(\vec{k}_{f_1}\vec{e}_{f_1})})_{1sn_1}(\gamma_{\mu_2}A_{\mu_2}^{*(\vec{k}_{f_2}\vec{e}_{f_2})})_{n_14s}}{E_{4s}-E_{n_1}-\omega}\right\rbrace\right|^2d\omega\;.
\end{eqnarray}
Interference contribution for the $ 4s-1s $ transition consists of 4 terms, corresponding to the four resonances, presented by Eqs. (\ref{69}), (\ref{72})
\begin{eqnarray}
\label{78}
dW^{2\gamma(interference)}_{4s-1s}(\omega)=\frac{2(\omega-\omega^{res1})F_1^{4s}(\omega)}{(\omega-\omega^{res1})^2+\frac{1}{4}(\Gamma_{4s}+\Gamma_{2p})^2}d\omega+\frac{2(\omega-\omega^{res2})F_2^{4s}(\omega)}{(\omega-\omega^{res2})^2+\frac{1}{4}\Gamma_{2p}^2}d\omega+\nonumber\\+\frac{2(\omega-\omega^{res3})F_3^{4s}(\omega)}{(\omega-\omega^{res3})^2+\frac{1}{4}(\Gamma_{4s}+\Gamma_{3p})^2}d\omega+\frac{2(\omega-\omega^{res4})F_4^{4s}(\omega)}{(\omega-\omega^{res4})^2+\frac{1}{4}\Gamma_{3p}^2}d\omega\;,
\end{eqnarray}
where 
\begin{eqnarray}
\label{79}
F^{4s}_i(\omega)=\sum\limits_{\vec{e}_{f_1}}\sum\limits_{\vec{e}_{f_2}}\int \frac{d\vec{\nu}_{f_1}}{(2\pi)^3}\frac{d\vec{\nu}_{f_2}}{(2\pi)^3}f^{4s}(\omega)\varphi_{i}^{4s},\; i=1,2,3,4\;,
\end{eqnarray}

\begin{eqnarray}
\label{80}
f^{4s}(\omega)=\omega^2(\omega^{4s}_0-\omega)^2\sum\limits_{n_1\neq 2p,3p}\left\lbrace \frac{(\gamma_{\mu_1}A_{\mu_1}^{*(\vec{k}_{f_2}\vec{e}_{f_2})})_{1sn_1}(\gamma_{\mu_2}A_{\mu_2}^{*(\vec{k}_{f_1}\vec{e}_{f_1})})_{n_14s}}{\omega+E_{1s}-E_{n_1}}+\right.\\\nonumber\left.+\frac{(\gamma_{\mu_1}A_{\mu_1}^{*(\vec{k_{f_1}}\vec{e_{f_1}})})_{1sn_1}(\gamma_{\mu_2}A_{\mu_2}^{*(\vec{k}_{f_2}\vec{e}_{f_2})})_{n_14s}}{E_{4s}-E_{n_1}-\omega}\right\rbrace\;,
\end{eqnarray}

\begin{eqnarray}
\label{81}
\varphi_{1}^{4s}=(\gamma_{\mu_1}A_{\mu_1}^{*(\vec{k}_{f_2}\vec{e}_{f_2})})_{1s2p}(\gamma_{\mu_2}A_{\mu_2}^{*(\vec{k}_{f_1}\vec{e}_{f_1})})_{2p4s}\;,
\end{eqnarray}

\begin{eqnarray}
\label{82}
\varphi_{2}^{4s}=(\gamma_{\mu_1}A_{\mu_1}^{*(\vec{k}_{f_1}\vec{e}_{f_1})})_{1s2p}(\gamma_{\mu_2}A_{\mu_2}^{*(\vec{k}_{f_2}\vec{e}_{f_2})})_{2p4s}\;,
\end{eqnarray}

\begin{eqnarray}
\label{81}
\varphi_{3}^{4s}=(\gamma_{\mu_1}A_{\mu_1}^{*(\vec{k}_{f_2}\vec{e}_{f_2})})_{1s3p}(\gamma_{\mu_2}A_{\mu_2}^{*(\vec{k}_{f_1}\vec{e}_{f_1})})_{3p4s}\;,
\end{eqnarray}

\begin{eqnarray}
\label{82}
\varphi_{4}^{4s}=(\gamma_{\mu_1}A_{\mu_1}^{*(\vec{k}_{f_1}\vec{e}_{f_1})})_{1s3p}(\gamma_{\mu_2}A_{\mu_2}^{*(\vec{k}_{f_2}\vec{e}_{f_2})})_{3p4s}\;.
\end{eqnarray}

\section{Three-photon $3p\rightarrow 1s+3\gamma $ transition}

In this section we consider the 3-photon transitions, taking as an example $3p\rightarrow 1s+3\gamma $ transition. The main decay channels of the $3p$ level are $ 3p\rightarrow 1s+\gamma $ and $ 3p\rightarrow 2s+\gamma $. Therefore as a resonance scattering process in this case we can choose the one-photon absorption and three-photon emission process depicted in Fig. 4. For the transition $3p\rightarrow 1s+3\gamma $ there are two cascades, containing two-photon links: $ 3p\rightarrow 2p+2\gamma\rightarrow 1s+3\gamma $ and $3p\rightarrow 2s+\gamma\rightarrow 1s+3\gamma$.

In case of the $ 3 $-photon transition we will take into account only cascade contribution. This cascade contribution contains necessarily one "pure two-photon" link ($ 3p-2p $ or $ 2s-1s $) and is, therefore of the same order of magnitude as the "pure two-photon" contribution to the $ 3s-1s $ transition. The  "pure 3-photon" transitions and the corresponding interference terms are essentially smaller than cascade contributions and can be neglected in the  "two-photon" approximation \cite{34}. In this sense the situation differs from the situation in $ 3s-1s $ two-photon decay when we were interested in the "pure two-photon" and interference terms.

The derivation similar to the two-photon case gives the following expression for the 3-photon emission amplitude $ 3p-1s $ in the resonance approximation:
\begin{eqnarray}
\label{83}
U^{em.3\gamma}_{3p-1s}=e^3\sum\limits_{n_1n_2}\frac{(\gamma_{\mu_1}A^{*\vec{k}_{f_3}\vec{e}_{f_3}}_{\mu_1})_{1sn_1}(\gamma_{\mu_2}A^{*\vec{k}_{f_2}\vec{e}_{f_2}}_{\mu_2})_{n_1n_2}(\gamma_{\mu_3}A^{*\vec{k}_{f_1}\vec{e}_{f_1}}_{\mu_3})_{n_23p}}{(E_{1s}-E_{n_1}+\omega_{f_3})(E_{1s}-E_{n_2}+\omega_{f_3}+\omega_{f_2})}\\\nonumber\times\frac{1}{E_{1s}-E_{3p}+\omega_{f_3}+\omega_{f_2}+\omega_{f_1}+\frac{i}{2}\Gamma_{3p}}\;.
\end{eqnarray}
In Eq. (\ref{83}) we summed already all the self-energy insertions in the lower electron propagator as is shown in Fig. 4. An exact energy conservation law in case of the process Fig. 4 is
\begin{eqnarray}
\label{83a}
\omega_i=\omega_{f_1}+\omega_{f_2}+\omega_{f_3}\;.
\end{eqnarray}
The resonance condition and the approximate energy conservation law in case of 3-photon decay looks like 
\begin{eqnarray}
\label{84}
|\omega_i-(E_{3p}-E_{1s})|=|\omega_{f_1}+\omega_{f_2}+\omega_{f_3}-(E_{3p}-E_{1s})|\leqslant \Gamma_{3p}\;.
\end{eqnarray}
To fix the cascade $3p\rightarrow 2s+\gamma \rightarrow1s+3\gamma  $ contribution  we set $ n_2=2s $ in Eq. (\ref{83}). This results 
\begin{eqnarray}
\label{85}
U^{em.3\gamma}_{3p-1s}=e^3\frac{1}{E_{1s}-E_{3p}+\omega_{f_3}+\omega_{f_2}+\omega_{f_1}+\frac{i}{2}\Gamma_{3p}}\frac{(\gamma_{\mu_1}A^{*\vec{k}_{f_1}\vec{e}_{f_1}}_{\mu_1})_{3p2s}}{(E_{1s}-E_{2s}+\omega_{f_3}+\omega_{f_2}+\frac{i}{2}\Gamma_{2s})}\times\\\nonumber\sum\limits_{n_1}\frac{(\gamma_{\mu_2}A^{*\vec{k}_{f_2}\vec{e}_{f_2}}_{\mu_2})_{2sn_1}(\gamma_{\mu_2}A^{*\vec{k}_{f_3}\vec{e}_{f_3}}_{\mu_3})_{n_11s}}{E_{1s}-E_{n_1}+\omega_{f_3}}+(perm)\;.
\end{eqnarray}
In Eq. (\ref{85}) we should include the contribution of the Feynman graphs with all the permutations of the photon lines (perm).

Now we have to take the right-hand side of Eq. (\ref{85}) by square modulus, to integrate over the emitted photon directions and to sum over the photon polarizations. Then we have to integrate over the photon frequencies $ \omega_{f_1} $, $ \omega_{f_2} $, $ \omega_{f_3} $ taking into account the condition Eq. (\ref{84}). However when we integrate the contribution of the cascade $ 3p-2s-1s $ we have to take into account that the frequency $ \omega_{f_1} $ is fixed by the resonance condition
\begin{equation}
\label{86}
| \omega_{f_1} -(E_{3p}-E_{2s})|\leqslant \Gamma_{3p}+\Gamma_{2s}\;.
\end{equation}
Inserting Eq. (\ref{86}) in the approximate conservation law (\ref{84}) we obtain the approximate equality
\begin{eqnarray}
\label{87}
\omega_{f_2}+\omega_{f_3}=E_{2s}-E_{1s}\;.
\end{eqnarray}
An integration over $ \omega_{f_2}$, $ \omega_{f_3} $ should be performed in the following way
\begin{eqnarray}
\label{88}
\int\limits_0^{E_{2s}-E_{1s}}d\omega_{f_2}\int\limits_0^{\omega_{f_2}}d\omega_{f_3}=\frac{1}{2}\int\limits_0^{E_{2s}-E_{1s}}d\omega_{f_2}\int\limits_0^{E_{2s}-E_{1s}}d\omega_{f_3}\;.
\end{eqnarray}
Eq. (\ref{88}) holds after symmetrization of Eq. (\ref{86}) via the permutation of the photons with the frequencies $ \omega_{f_2} $, $ \omega_{f_3} $.
The integration over two frequencies (e.g. over $ \omega_{f_1} $ and $ \omega_{f_2} $ in Eq. (\ref{85})) can be always extended to the interval  [$-\infty$, $+\infty$] since the pole approximation can be used. The third integration over $ \omega_{f_3} $ in Eq. (\ref{85}) according to Eq. (\ref{88}) should be performed over the finite interval [0, $E_{2s}-E_{1s}$]. The integration yields  
\begin{eqnarray}
\label{89}
b^{3\gamma}_{3p-1s}(3p-2s-1s)=2\pi e^3 \frac{\Gamma_{3p}+\Gamma_{2s}}{\Gamma_{3p}\Gamma_{2s}}\int\limits_{-\infty}^{\infty}\omega^2_{_{f_1}}d\omega_{f_1}\sum\limits_{\vec{e}_{f_1}}\int \frac{d\vec{\nu}_{f_1}}{(2\pi)^3}\frac{|U^{em.1\gamma}(3p-2s-1s)|}{(E_{3p}-E_{2s}-\omega_{f_1})^2+\frac{1}{4}(\Gamma_{3p}+\Gamma_{2s})^2}\times\\\nonumber\frac{1}{2}\int\limits_0^{\omega_{max}}\omega^2_{f_3}(\omega_{max}-\omega_{f_3})^2d\omega_{f_3}\sum\limits_{\vec{e}_{f_2}}\sum\limits_{\vec{e}_{f_3}}\frac{d\vec{\nu}_{f_2}}{(2\pi)^3}\frac{d\vec{\nu}_{f_3}}{(2\pi)^3}\sum\limits_{n_1}\frac{(\gamma_{\mu_2}A_{\mu_2}^{*(\vec{k}_{f_3}\vec{e}_{f_3})})_{1sn_1}(\gamma_{\mu_3}A_{\mu_3}^{*(\vec{k}_{f_2}\vec{e}_{f_2})})_{n_12s}}{E_{1s}-E_{n_1}+\omega_{f_3}}=\frac{W^{1\gamma}_{3p-2s}W^{2\gamma}_{2s-1s}}{\Gamma_{3p}\Gamma_{2s}}\;,
\end{eqnarray}
where $ \omega_{max}=E_{2s}-E_{1s} $.

The physical sense of the dimensionless quantity $ b^{3\gamma}_{3p-1s}(3p-2s-1s) $ should be discussed specially. This quantity should define the $ 3\gamma $ transition rate $ 3p-1s $
 via the channel $ 3p\rightarrow2s+\gamma \rightarrow 1s+3\gamma $. This transition rate is very small compared to the main decay channel for the $ 3p $ state, i.e. $ W^{1\gamma}_{3p-1s} $. We assume that the quantity $ b^{3\gamma}_{3p-1s}(3p-2s-1s) $ is the branching ratio for the $ 3 $-photon transition rate $ W^{3\gamma}_{3p-1s}(3p-2s-1s) $ to the direct two-photon transition rate $ W^{2\gamma}_{2s-1s}=\Gamma_{2s} $. Then from Eq. (\ref{89}) it follows
\begin{eqnarray}
\label{90}
W^{3\gamma}_{3p-1s}(3p-2s-1s)=\frac{W^{1\gamma}_{3p-2s}}{\Gamma_{3p}}W^{2\gamma}_{2s-1s}\;.
\end{eqnarray}
In the same way the contribution of the $ 3 $-photon cascade $ 3p-2p+2\gamma\rightarrow 1s+3\gamma $ can be analysed. The final result looks like 
\begin{eqnarray}
\label{91}
b^{3\gamma}(3p-2p-1s)=\frac{W^{1\gamma}_{2p-1s}}{\Gamma_{2p}\Gamma_{3p}}W^{2\gamma}_{3p-2p}\;.
\end{eqnarray}
Unlike $b^{3\gamma}(3p-2s-1s)$ the quantity Eq. (\ref{91}) should be considered as the branching ratio of the transition rate via channel $ 3p-2p-1s $ to the total width of the $ 3p $ level, i.e. $ \Gamma_{3p} $. Then
\begin{eqnarray}
\label{92}
W^{3\gamma}_{3p-1s}(3p-2p-1s)=\frac{W^{1\gamma}_{2p-1s}}{\Gamma_{2p}}W^{2\gamma}_{3p-2p}\;.
\end{eqnarray}
In the equation in \cite{35}, corresponding to the Eq. (\ref{92}), the factor $ W^{1\gamma}_{2p-1s}/\Gamma_{2p}=1$ was omitted. Here we keep it to demonstrate that the transition channel $ 3p-2p-1s $ is a $ 3 $-photon channel. Total expression for the transition rate $ 3p-1s $ ($ 1 $-photon and $ 3 $-photon) is
\begin{eqnarray}
\label{93}
W^{3\gamma}_{3p-1s}=W^{1\gamma}_{3p,1s}+\frac{W^{1\gamma}_{2p,1s}}{\Gamma_{2p}}W^{2\gamma}_{3p,2p}+\frac{W^{1\gamma}_{3p,2s}}{\Gamma_{3p}}W^{2\gamma}_{2s,1s}.
\end{eqnarray}
This expression coincides with one derived in \cite{35} up to the coefficients before the second and third terms in the right-hand side of Eq. (\ref{93}). In \cite{35} this coefficients were evaluated incorrectly and were equal $ 3/4 $. 

\section{QM approach for the two-photon decays}
 
In this section we will follow along the lines of the derivation for the one-photon transition given in \cite{40}. The QM approach is based on the time dependent Schr\"{o}dinger equation
\begin{eqnarray}
\label{95}
i\frac{\partial\psi(t)}{\partial t}=(\widehat{H}_0+\widehat{V}(t))\psi(t),
\end{eqnarray}
where $ \widehat{H}_0 $ is a time-independent zero-order atomic Hamiltonian and $ \widehat{V}(t) $ is the perturbation describing the interaction of an atom with the photon field. According to the the Dirac perturbation theory we search for the solution of Eq. (\ref{95}) in the form of the expansion 
\begin{eqnarray}
\label{96}
\psi(t)=\sum\limits_{\nu}a_{\nu}(t)e^{-i\varepsilon_{\nu}t}\psi^{(0)}_{\nu},
\end{eqnarray}
where $ \psi^{(0)}_{\nu} $ are the eigenfunctions of $ \widehat{H}_0 $:
\begin{eqnarray}
\label{97}
\widehat{H}_0\psi^{(0)}_{\nu}=E_{\nu}\psi^{(0)}_{\nu}
\end{eqnarray}
and $ \epsilon_{\nu} $ are the energies of the system atom + photon field. These energies can be presented as
\begin{eqnarray}
\label{98}
\varepsilon_{\nu}=E_{\nu}-\frac{i}{2}\Gamma_{\nu}+\sum\omega,
\end{eqnarray}
where $ E_{\nu} $ re the eigenstates of the Hamiltonian $ \widehat{H}_0 $, $ \Gamma_{\nu} $ are the corresponding level widths and $ \sum\omega $ is the sum of the photons which arrives in the field, when an atom arrives in the state $ \nu $.
In Eq.(\ref{96}) we need also a second subscript, indicating to which stationary state $ \psi^{(0)}_{\mu} $ we apply the perturbation theory:
\begin{eqnarray}
\label{99}
\psi_{\mu}(t)=\sum\limits_{\nu}a_{\mu\nu}(t)e^{-i\varepsilon_{\nu}t}\psi^{(0)}_{\nu},
\end{eqnarray}
For the coefficients $ a_{\mu\nu}(t) $ from Eq. (\ref{95}) follows the system of equations
\begin{eqnarray}
\label{100}
i\frac{\partial a_{\mu\nu}(t)}{\partial t}=\sum\limits_{\nu'}\widehat{V}_{\nu\nu'}a_{\mu\nu}(t)e^{i(\varepsilon_{\nu}-\varepsilon_{\nu'})t},
\end{eqnarray}
where $ \widehat{V}_{\nu\nu'} $ are the matrix elements of the operator $ \widehat{V} $. In these matrix elements we understand this operator as time-independent, the time dependence of it being included in the exponent $ e^{-i\varepsilon_{\nu} t} $.
We are interested in evaluation of quantities
\begin{eqnarray}
\label{101}
\lim\limits_{t\rightarrow\infty}|a_{\mu\nu}(t)|^2=W_{\mu\nu}\;,
\end{eqnarray}
which can be interpreted as the transition probabilities from the state $ \mu $ to the state $ \nu $ of the atom+field system.
To derive the two-photon transition probability we need to employ the second-order perturbation theory. In the zero-order approximation for initial state $ \mu=i $ we have 
\begin{eqnarray}
\label{102}
a^{(0)}_{i\nu}=\delta_{i\nu},
\end{eqnarray}
where $ \delta_{i\nu} $ is the Kronecker symbol. This zero-order value should be inserted then in the right-hand side of Eq. (\ref{100}). Then the equation for the first-order correction $ a^{(1)}_{i\nu} $ results

\begin{eqnarray}
\label{103}
i\frac{\partial a^{(1)}_{i\nu}}{\partial t}=\widehat{V}_{\nu i}e^{i(\varepsilon_{\nu}-\varepsilon_i)t}.
\end{eqnarray}
Integrating Eq. (\ref{103}) we find
\begin{eqnarray}
\label{104}
a_{i\nu}^{(1)}=\frac{\widehat{V}_{\nu i}(1-e^{i(\varepsilon_{\nu}-\varepsilon_i)t})}{\varepsilon_{\nu}-\varepsilon_i}.
\end{eqnarray}
Here we have to understand
\begin{eqnarray}
\label{105}
\varepsilon_i =E_i-\frac{i}{2}\Gamma_i\;,
\end{eqnarray}
\begin{eqnarray}
\label{106}
\varepsilon_{\nu} =E_{\nu}-\frac{i}{2}\Gamma_{\nu}+\omega \;,
\end{eqnarray}
where $ \nu $ is the intermediate state in two-photon transition and $ \omega $ is the frequency of the photon emitted in the transition $ i\rightarrow \nu $. 
As the next step we have to insert Eq. (\ref{104}) into the right-hand side of Eq. (\ref{100}) and to set $ \nu=f $ (final state). This yields
\begin{eqnarray}
\label{107}
i\frac{\partial a^{(2)}_{if}(t)}{\partial t}=\sum\limits_{\nu}\frac{\widehat{V}_{f\nu}\widehat{V}_{\nu i}}{\varepsilon_i-\varepsilon_{\nu}}[1-e^{i(\varepsilon_{\nu}-\varepsilon_i)t}]e^{i(\varepsilon_{f}-\varepsilon_{\nu})t}.
\end{eqnarray}
Here

\begin{eqnarray}
\label{108}
\varepsilon_f =E_f-\frac{i}{2}\Gamma_f +\omega + \omega'
\end{eqnarray}
with the condition
\begin{eqnarray}
\label{109}
\omega_1 + \omega_2 = E_i-E_f.
\end{eqnarray}
If $ f $ is the ground state $ f=0 $, $ \Gamma_0=0 $.

Integrating Eq. (\ref{107}) over time we find second-order correction
\begin{eqnarray}
\label{110}
a^{(2)}_{if}=\sum\limits_{\nu'}\frac{\widehat{V}_{f\nu'}\widehat{V}_{\nu i}}{\varepsilon_{\nu'}-\varepsilon_i}\left(\frac{1-e^{i(\varepsilon_{f}-\varepsilon_{\nu'})t}}{\varepsilon_{f}-\varepsilon_{\nu'}}-\frac{1-e^{i(\varepsilon_{f}-\varepsilon_{i})t}}{\varepsilon_{f}-\varepsilon_{i}}\right),
\end{eqnarray}
In Eq. (\ref{110}) we should set $  t\rightarrow \infty $ according to definition (\ref{101}). Taking into account permutation symmetry we find finally equation for transition probability
\begin{eqnarray}
\label{111}
dW_{if}^{(2)}=\lim\limits_{t\rightarrow\infty}|U_{if}^{(2)}|^2d\omega d\omega'\;,
\end{eqnarray}
where $ U_{if}^{(2)}=a^{(2)}_{if}+a^{(2)}_{if}(\omega\leftrightarrow\omega') $. A simple algebraic transformation yields
\begin{eqnarray}
\label{112}
U_{if}^{(2)}=\left\lbrace\sum\limits_{\nu}\frac{\widehat{V}_{f\nu'}\widehat{V}_{\nu i}}{(E_f-E_{\nu} +\omega +\frac{i}{2}(\Gamma_{\nu}-\Gamma_f))}+ \sum\limits_{\nu}\frac{\widehat{V}_{f\nu}\widehat{V}_{\nu i}}{(E_f-E_{\nu} +\omega +\frac{i}{2}(\Gamma_{\nu}-\Gamma_f))}\right\rbrace\times\\\nonumber\times\frac{1}{E_f-E_i +\omega+\omega'+\frac{i}{2}(\Gamma_i-\Gamma_f)}.
\end{eqnarray}
We see that Eq. (\ref{112}) coincides with QED Eq. (\ref{53}) in case of $ i=3s $, $ f=1s $, $ \nu=2p $, $ \omega=\omega_{f_1} $ and $ \omega'=\omega_{f_2} $. The further derivations are exactly the same as in sections IV, V and the final results are also exactly the same. Similar considerations can be used in case of 3-photon transitions. For the derivation of the 3-photon transition probability we have to go to the third order of the perturbation theory.

Substituting Eq. (\ref{110}) into right-hand side of Eq. (\ref{100}) for the third-order correction $ a^{(3)}_{i\nu} $ we arrive at

\begin{eqnarray}
\label{113}
i\frac{\partial a^{(3)}_{if}}{\partial t}=\sum\limits_{n\nu}\frac{V_{fn}V_{n\nu}V_{\nu i}}{\varepsilon_{\nu}-\varepsilon_{i}}\left(\frac{1-e^{i(\varepsilon_n-\varepsilon_{\nu})t}}{\varepsilon_n-\varepsilon_{\nu}}-\frac{1-e^{i(\varepsilon_n -\varepsilon_i)t}}{\varepsilon_n-\varepsilon_i}\right)e^{i(\varepsilon_f -\varepsilon_n)t}\;,
\end{eqnarray}
Here
\begin{eqnarray}
\label{114}
\varepsilon_n =E_n-\frac{i}{2}\Gamma_n +\omega + \omega'\;,
\end{eqnarray}
\begin{eqnarray}
\label{115}
\varepsilon_f =E_f-\frac{i}{2}\Gamma_f +\omega + \omega' +\omega''\;.
\end{eqnarray}
Taking the limit $ t\rightarrow\infty $ we can find from Eq. (\ref{113})
\begin{eqnarray}
\label{115a}
 a^{(3)}_{if}=\sum\limits_{n\nu}\frac{V_{fn}V_{n\nu}V_{\nu i}}{(\varepsilon_f-\varepsilon_i)(\varepsilon_f-\varepsilon_n)(\varepsilon_f-\varepsilon_{\nu})}\;.
\end{eqnarray}
Substituting Eqs. (\ref{105}), (\ref{106}), (\ref{114}) and (\ref{115}) into Eq. (\ref{115a}) yields
\begin{eqnarray}
\label{116}
 a^{(3)}_{if}=\sum\limits_{n\nu}\frac{V_{fn}V_{n\nu}V_{\nu i}}{(E_f-E_n  +\omega''+\frac{i}{2}(\Gamma_n-\Gamma_f))(E_f-E_{\nu}+ \omega' +\omega''+\frac{i}{2}(\Gamma_{\nu}-\Gamma_f )}\times\\\nonumber\times\frac{1}{(E_f-E_i+\omega + \omega' +\omega''+\frac{i}{2}(\Gamma_i-\Gamma_f))}\;.
\end{eqnarray}
Eq. (\ref{116}) coincides with QED equation (\ref{83}) in case of $ i=3p $, $ f=1s $. The further derivations will apparently give the same result as obtained with QED approach in section VII.
\section{Comparison of the different regularizations of the cascade contributions}
To estimate the importance of the regularization of the type Eq. (\ref{4}) with two widths in the resonance denominator $ \Gamma_i+\Gamma_{int} $ where $ \Gamma_i $ is the initial state and $ \Gamma_{int} $ is the intermediate state for the cascade $ i\rightarrow int+\gamma\rightarrow f+ 2\gamma $ ($ f $ denotes the final state) not only at the point of the resonance but also in the wings of the resonance, we introduce the quantity
\begin{eqnarray}
\label{125a}
\eta=\frac{\Gamma_i+\Gamma_{int}}{\Delta}\times\frac{\Gamma_i}{\Gamma_{int}+\Gamma_i}\;,
\end{eqnarray}
where $ \Delta $ is the distance from the point of the resonance. The first factor in Eq. (\ref{125a}) characterizes the influence of the resonance regularization at the distance $ \Delta $ in the wing and the second factor defines the relative importance of the regularization with two widths. It is convenient to express the distance $ \Delta $ in terms of the resonance width, i.e. to choose 
\begin{eqnarray}
\label{125b}
\Delta= m(\Gamma_i+\Gamma_{int})
\end{eqnarray}
where $ m $ is some number. Then
\begin{eqnarray}
\eta=\frac{1}{m}\frac{\Gamma_i}{\Gamma_i+\Gamma_{int}}\;.
\end{eqnarray}
The $ \eta $ values are given in Table I for transitions $ 3s-1s $ and $ 3d-1s $ and the $ m $ values up to $ m=100 $. 

To present more special illustration of the proper choice of the cascade regularization we will evaluate an absorbability of the two-photon emission presented by the frequency distribution Eq. (\ref{2}) for the $ 3s-1s $ two-photon transition. Similar property was first introduced in \cite{33} and employed in \cite{48}, \cite{49} to characterize the probability of absorption of the radiation emitted by one atom and absorbed by another atom. In our case this means the probability for the two-photon radiation emitted by one atom to be absorbed in the one-photon absorption lines of another atom.

It is convenient to begin with the absorbability $ X^{2\gamma}_{2s-1s} $ of two-photon emission, to compare with this quantity the absorbability of the other two-photon transitions. We define $ X^{2\gamma} $ as \cite{49}
\begin{eqnarray}
\label{117}
X^{2\gamma}_{2s-1s}=\frac{1}{2}\int\limits_0^{\infty}L^{1\gamma}_{2p-1s}(\omega)dW^{2\gamma}_{2s-1s}(\omega)\;,
\end{eqnarray}
where $ dW^{2\gamma}_{2s-1s}(\omega) $ is the frequency distribution (differential transition rate) for $ 2s-1s $ two-photon transition and $ L^{2\gamma}_{2p-1s} $ is the Lorentz line profile for the one-photon $ 1s-2p $ absorption line, which is the same as the Lorentz profile for the $ 2p-1s $ emission line. The Lorentz profile is given by Eq. (\ref{31}). A dimensionless frequency distribution $ \frac{dW^{2\gamma}_{2s-1s}}{d\omega} $ is normalized as
\begin{eqnarray}
\label{118}
\frac{1}{2}\int\limits_0^{\omega_{2s-1s}}dW^{2\gamma}_{2s-1s}(\omega)=W^{2\gamma}_{2s-1s}=8.229\;s^{-1}\;.
\end{eqnarray}
In the integral Eq. (\ref{118}) outside the interval $ [0,\omega_{2s-1s}] $ where $ \omega_{2s-1s}=E_{2s}-E_{1s} $ the function $ dW^{2\gamma}_{2s-1s}(\omega) $ should be set to zero. Actually, the normalization conditions Eq. (\ref{32}) and Eq. (\ref{118}) are not important since only comparison of $ X^{2\gamma}_{2s-1s} $ with the corresponding quantities for the other two-photon transitions matters. Anyhow, a dimensionless quantity $ X^{2\gamma}_{2s-1s} $ can be considered as an absorbability, i.e. an absolute probability of the absorption of the $ 2s-1s $ two-photon emission by the Lyman-alpha one-photon absorption line. A numerical evaluation gives the result
\begin{eqnarray}
\label{119}
X^{2\gamma}_{2s-1s}=6.50\times10^{-22}\;,
\end{eqnarray}
which shows that the two-photon radiation $ 2s-1s $ emitted by one hydrogen atom can not be absorbed via Lyman-alpha transition by another hydrogen atom in the ground state. Note that the transition $ 2s-1s $ does not contain cascades, i.e. the frequency distribution $ dW^{2\gamma}_{2s-1s} $ has no singularities which have to be regularized. The quantity
\begin{eqnarray}
\label{120}
Y^{2\gamma}_{2s-1s}=1-X^{2\gamma}_{2s-1s}
\end{eqnarray}
can be called "nonabsorbability"\; and for $ 2s-1s $ two-photon transition
\begin{eqnarray}
\label{121}
Y^{2\gamma}_{2s-1s}=1\;.
\end{eqnarray}
In Table II we present the results of calculations for the absorbabilities for the two-photon $ 3s-1s $, $ 3d-1s $ and $ 4d-1s $ transitions. All these transitions require the regularization of cascade contributions. The emission of $ 3s-1s $ and $ 3d-1s $ transitions can be absorbed by several one-photon absorption lines: $ 1s-2p\;,2s-3p\;,2p-3s\;,2p-3d\ $. As it was explained in the Introduction we consider only $ E1 $ transitions in hydrogen atom. Then the generalization of Eq. (\ref{117}) to the $ 3s(3d)-1s $ two-photon transition should be
\begin{eqnarray}
\label{122}
X^{2\gamma}_{3s(3d)-1s}=\frac{1}{2}\int\limits_0^{\infty}\left[L^{1\gamma}_{2p-1s}(\omega)+L^{1\gamma}_{3p-2s}(\omega)+L^{1\gamma}_{3s-2p}(\omega)+L^{1\gamma}_{3d-2p}(\omega)\right]dW^{2\gamma}_{3s(3d)-1s}(\omega)\;,
\end{eqnarray}
Note, that the Lorentz line profile for the transition between to excited states $ nl $ and $ n'l' $ Eq. (\ref{31}) is given by an expression \cite{40}
\begin{eqnarray}
\label{123}
L^{1\gamma}_{nl\rightarrow n'l'}(\omega)=\frac{1}{2\pi}\frac{W^{1\gamma}_{nl\rightarrow n'l'}}{(\omega+E_{n'l'}-E_{nl})^2+\frac{1}{4}(\Gamma_{nl}+\Gamma_{n'l'})^2}\;.
\end{eqnarray}
The Lamb shifts of the levels $ nl $, $ n'l' $ are neglected. In a similar way the quantity $ X^{2\gamma}_{4d-1s} $ can be constructed.

In Table II the two values for $ X^{2\gamma}_{3s-1s} $ are given: first, the value $ X^{2\gamma(1)}_{3s-1s} $ evaluated with the frequency distribution $ dW^{2\gamma}_{3s-1s}(\omega) $ where the upper link of the cascade $ 3s-2p-1s $ is regularized according to Eq. (\ref{4}) and second, the value $ X^{2\gamma(2)}_{3s-1s} $ obtained with the regularization via Eq. (\ref{9}). The same is done for $ X^{2\gamma}_{3d-1s} $; the regularization formulas for this case can be obtained from Eq. (\ref{4}) and Eq. (\ref{9}) by replacement $ 3s\rightarrow 3d $. A similar calculation is done for $ X^{2\gamma}_{4d-1s} $, though the cascade links in this case are different from cascade links of $ 3s(3d)-1s $ transitions. In the Table II the standard values for $ \Gamma_{2s}=W^{2\gamma}_{2s-1s} $, $ \Gamma^{1\gamma}_{3s}=W^{1\gamma}_{3s-2p} $ , $ \Gamma^{1\gamma}_{3d}=W^{1\gamma}_{3d-2p} $, $ \Gamma^{1\gamma}_{4d}=W^{1\gamma}_{4d-3p}+W^{1\gamma}_{4d-2p} $ are also indicated. A difference between $ X^{2\gamma(1)}_{3s-1s} $ and $ X^{2\gamma(2)}_{3s-1s} $ as well as between $ X^{2\gamma(1)}_{3d-1s} $ and $ X^{2\gamma(2)}_{3d-1s} $ demonstrates the relative importance of employing the proper regularization scheme. In case of $ 3d-1s $ transition the deviation becomes more than $ 10\% $ of the total $ X^{2\gamma} $ value.

\section{Conclusions}
In this paper we analyzed the problem of the multiphoton transitions with cascades taking as an example the two-photon $  3s\rightarrow 1s+2\gamma$, $ 4s\rightarrow 1s+2\gamma $ transitions and the three-photon $ 3p\rightarrow 1s+3\gamma $ transition. We proved that the regularization of the singularities in the expressions for the cascade contributions to the transition rates should include the widths of both initial and intermediate states.

The analysis of the problem was performed within QED approach which allows, in principle, to describe any multiphoton process with any number of cascades. We demonstrated also that the same results can be obtained within QM phenomenological approach provided that one follows the proper evaluation scheme based on the time-dependent Schr\"{o}dinger equation. Few examples are given to demonstrate the importance of the correct regularization of the cascade resonances, especially when the multiphoton frequency distributions are converted with some other photon distribution. 

\section*{Acknowledgements}
The authors wish to thank Dr. O. Yu. Andreev for many helpful discussions. This work was supported by RFBR grant 11-02-00168a and by The Ministry of education and science of Russian Federation, project 8420. T. Z. acknowledges support by the non-profit Foundation "Dynasty" (Moscow). 

\setcounter{equation}{0}
\renewcommand{\theequation}%
{A.\arabic{equation}}

\section*{Appendix A: Derivation of Eqs. (\ref{54}), (\ref{58})}
Taking the first term in the curly brackets in Eq. (\ref{53}) together with the factor outside the brackets by square modulus, integrating over the emitted photons directions and introducing the shorthand notations
\begin{eqnarray}
E_{2p}-E_{1s}=\omega^{res. 2}\equiv \Delta E_A\;,
\end{eqnarray}
\begin{eqnarray}
E_{3s}-E_{1s}\equiv\omega_0\equiv \Delta E_B\;,
\end{eqnarray}
we define the double differential branching ratio as 
\begin{eqnarray}
db^{2\gamma(resonance\;1)}_{3s-2p-1s}=\frac{1}{(2\pi)^2}\frac{W^{1\gamma}_{3s-2p}(\omega^{res.1})W^{1\gamma}_{2p-1s}(\omega^{res.2})}{[\Delta E_A-\omega_{f_2}-\frac{i}{2}\Gamma_{2p}][\Delta E_A-\omega_{f_2}+\frac{i}{2}\Gamma_{2p}]}\times\\\nonumber\frac{d\omega_{f_1}d\omega_{f_2}}{[\Delta E_B-\omega_{f_1}-\omega_{f_2}-\frac{i}{2}\Gamma_{3s}][\Delta E_B-\omega_{f_1}-\omega_{f_2}+\frac{i}{2}\Gamma_{3s}]}\;.
\end{eqnarray}
Using Cauchy theorem we integrate over $ \omega_{f_2} $ in the lower half-plane where the poles are:
\begin{eqnarray}
\omega^{(1)}_{f_2}=\Delta E_A-\frac{i}{2}\Gamma_{2p}\;,
\end{eqnarray}
\begin{eqnarray}
\omega^{(2)}_{f_2}=\Delta E_B-\omega_{f_1}-\frac{i}{2}\Gamma_{3s}\;.
\end{eqnarray}
The integration results:

\begin{eqnarray}
db^{2\gamma(resonance\;1)}_{3s-2p-1s}=\frac{1}{2\pi}W^{1\gamma}_{3s-2p}(\omega^{res.1})W^{1\gamma}_{2p-1s}(\omega^{res.2})\times\\\nonumber\left(\frac{1}{\Gamma_{2p}[\Delta E_B-\Delta E_A-\omega_{f_1}+\frac{i}{2}(\Gamma_{2p}-\Gamma_{3s})][\Delta E_B-\Delta E_A-\omega_{f_1}+\frac{i}{2}(\Gamma_{2p}+\Gamma_{3s})]}\right.+\\\nonumber\left.+\frac{1}{\Gamma_{3s}[\Delta E_A-\Delta E_B+\omega_{f_1}-\frac{i}{2}(\Gamma_{2p}-\Gamma_{3s})][\Delta E_A-\Delta E_B+\omega_{f_1}+\frac{i}{2}(\Gamma_{2p}+\Gamma_{3s})]}\right)  d\omega_1\;.
\end{eqnarray}
Algebraic transformations then lead to
\begin{eqnarray}
db^{2\gamma(resonance\;1)}_{3s-2p-1s}=\frac{1}{2\pi}\frac{W^{1\gamma}_{3s-2p}(\omega^{res.1})W^{1\gamma}_{2p-1s}(\omega^{res.2})}{\Gamma_{2p}\Gamma_{3s}}\times\\\nonumber\frac{\Gamma_{3s}[\Delta E_B-\Delta E_A-\omega_{f_1}-\frac{i}{2}(\Gamma_{2p}+\Gamma_{3s})]+\Gamma_{2p}[\Delta E_B-\Delta E_A-\omega_{f_1}+\frac{i}{2}(\Gamma_{2p}+\Gamma_{3s})]}{[ \Delta E_B-\Delta E_A-\omega_{f_1}+\frac{i}{2}(\Gamma_{2p}-\Gamma_{3s})][(\Delta E_B-\Delta E_A-\omega_{f_1})^2+\frac{1}{4}(\Gamma_{2p}+\Gamma_{3s})^2]}=\\\nonumber=\frac{1}{2\pi}\frac{W^{1\gamma}_{3s-2p}(\omega^{res.1})W^{1\gamma}_{2p-1s}(\omega^{res.2})}{\Gamma_{2p}\Gamma_{3s}}\times\\\nonumber\frac{(\Gamma_{3s}+\Gamma_{2p})[\Delta E_B-\Delta E_A-\omega_{f_1}+\frac{i}{2}(\Gamma_{2p}-\Gamma_{3s})]}{[\Delta E_B-\Delta E_A-\omega_{f_1}+\frac{i}{2}(\Gamma_{2p}-\Gamma_{3s})][(\Delta E_B-\Delta E_A-\omega_{f_1})^2+\frac{1}{4}(\Gamma_{2p}+\Gamma_{3s})^2]}\;.
\end{eqnarray}
After the cancellation of the factor $ [\Delta E_B-\Delta E_A-\omega_{f_1}+\frac{i}{2}(\Gamma_{2p}-\Gamma_{3s})] $ in the numerator and the denominator of Eq. (A.7) we arrive of the expression (\ref{54}) given in the text.

To obtain Eq. (\ref{58}) we have to take the second term in the curly brackets in Eq. (\ref{53}), to integrate it over the directions of the emitted photons and to sum over the emitted photons polarizations. This gives
\begin{eqnarray}
db^{2\gamma(resonance\;2)}_{3s-2p-1s}=\frac{1}{2\pi}\frac{W^{1\gamma}_{3s-2p}(\omega^{res.1})W^{1\gamma}_{2p-1s}(\omega^{res.2})d\omega_{f_1}d\omega_{f_2}}{[(\Delta E_A-\omega_{f_1})^2+\frac{1}{4}\Gamma^2_{2p}][(\Delta E_B-\omega_{f_1}-\omega_{f_2})^2+\frac{1}{4}\Gamma^2_{3s}]}\;.
\end{eqnarray}
We again integrate over $ \omega_{f_2} $ in the complex plane obtaining immediately the result Eq. (\ref{58}).

\begin{table}[hP]
\caption{Relative importance $ \eta $ of the cascade regularization with two widths at the different distances from the point of the resonance}
\begin{tabular}{| c | c | c | c | c | c |}
\cline{4-6}
\multicolumn{1}{c}{}&\multicolumn{1}{c}{} &\multicolumn{1}{c|}{} &\multicolumn{3}{c|}{$  m $}\\
\hline $ transition $ & $ \Gamma_i,\;s^{-1} $ & $ \Gamma_{int},\;s^{-1} $ & $ 1 $  & $ 10 $ & $ 100 $
\qquad\\
\hline $ 3s-1s $ & $ \Gamma_{3s}=0.063170\times 10^8 $ & $ \Gamma_{2p}=6.268258\times 10^8 $ & $ 0.01  $  & $ 0.001 $ & $ 0.0001 $ 
\qquad\\
\hline $ 3d-1s $ & $ \Gamma_{3d}=0.646857\times10^8 $ & $ \Gamma_{2p}=6.268258\times 10^8  $ &  $ 0.1  $  & $ 0.01 $ & $ 0.001 $ 
\qquad\\
\hline
\end{tabular}
\end{table}

\begin{table}[hP]
\caption{Absorbability for the two-photon emission $ nl\rightarrow1s+2\gamma $ with different cascade regularizations}
\begin{tabular}{| l | c | c | c | c | c | c | r |}
\hline\hline
$ nl$ & $ X^{2\gamma(1)}_{nl-1s} $ & $ X^{2\gamma(2)}_{nl-1s} $ & $ Y^{2\gamma(1)}_{nl-1s} $  & $ Y^{2\gamma(2)}_{nl-1s} $ & $ \Gamma_{nl}=W^{2\gamma}_{nl-1s}\;,s^{-1} $
\qquad\\
\hline $ 2s $ & $ 6.393\times10^{-22} $ & $ 6.393\times10^{-22} $ & $ 1.00000  $  & $ 1.00000 $ & $ 8.22935 $ 
\qquad\\
\hline $ 3s $ & $ 0.00497 $ & $ 0.00502  $ & $ 0.99504  $  & $ 0.99498 $ & $ 0.06317\times10^{8} $ 
\qquad\\
\hline $ 3d $ & $ 0.04652 $ & $ 0.05217  $ & $ 0.95349  $  & $ 0.94783 $ & $ 0.64686\times10^{8} $ 
\qquad\\
\hline $ 4d $ & $ 0.02118 $ & $ 0.02385 $ & $ 0.97882  $  & $ 0.97615 $ & $ 0.26013\times10^{8} $ 
\qquad\\
\hline
\end{tabular}
\end{table}

\newpage
\begin{figure}[h!]
  \centering
\includegraphics[scale=0.4]{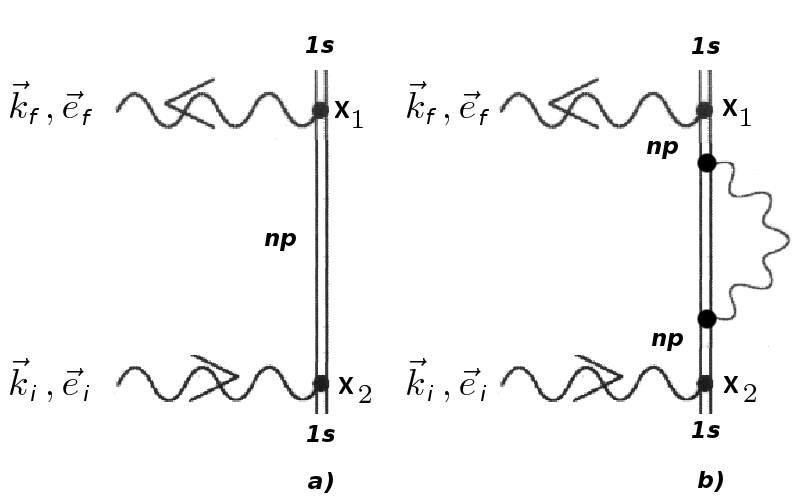}
  \caption{Feynman graph describing the resonant photon scattering on the ground state of hydrogen atom. In FIG. 1$ a $ the basic process of the resonant scattering with the excitation of $ np $ state is depicted. In FIG. 1$ b $ the electron self-energy insertion in the propagator is made. The double solid lines denote the electron in the field of the nucleus (Furry picture of QED), the wavy lines denote the absorbed, emitted and virtual photons.}
\end{figure}
\begin{figure}[h!]
  \centering
\includegraphics[scale=0.4]{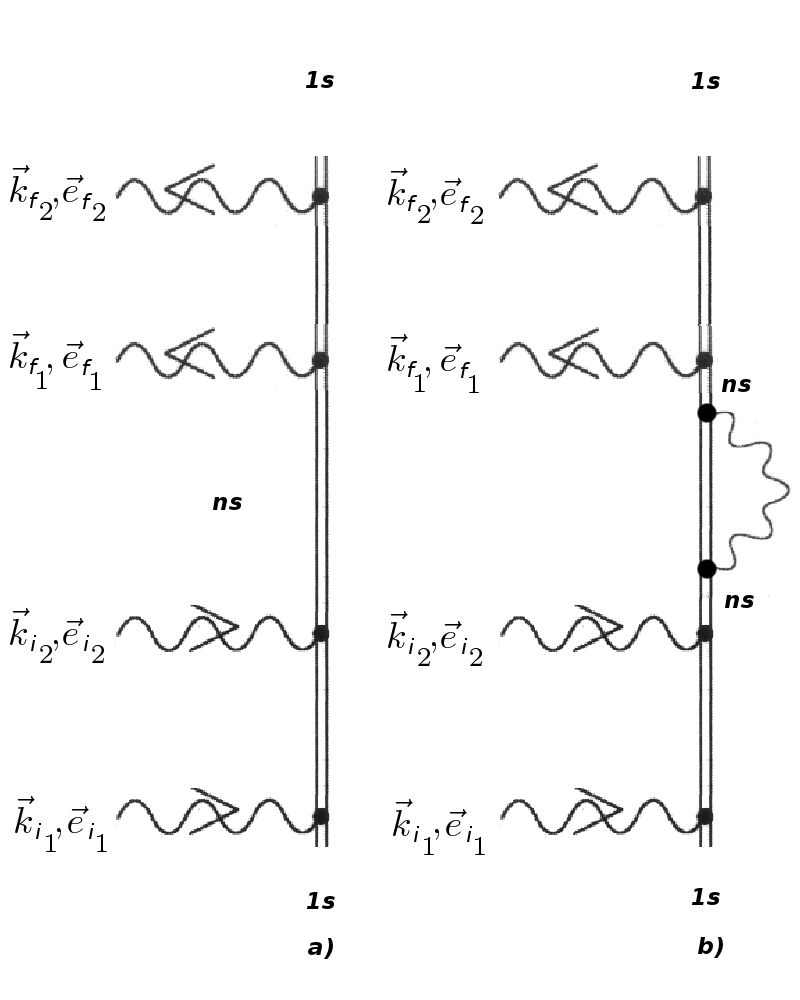}
  \caption{Feynman graph describing the two-photon resonant scattering on the ground state of hydrogen atom, with the excitation of $ ns $ $ (n>2) $ state and the resonance condition $ \omega_1+\omega_2=E_{ns}-E_{1s} $. In FIG. 2$ a $ the basic process of the resonant scattering with the excitation of the $ ns $ state is depicted. In FIG. 2$ b $ the electron self-energy insertion in the central electron propagator is made. The notations are the same as in FIG. 1}
\end{figure}
\begin{figure}[h!]
  \centering
\includegraphics[scale=0.4]{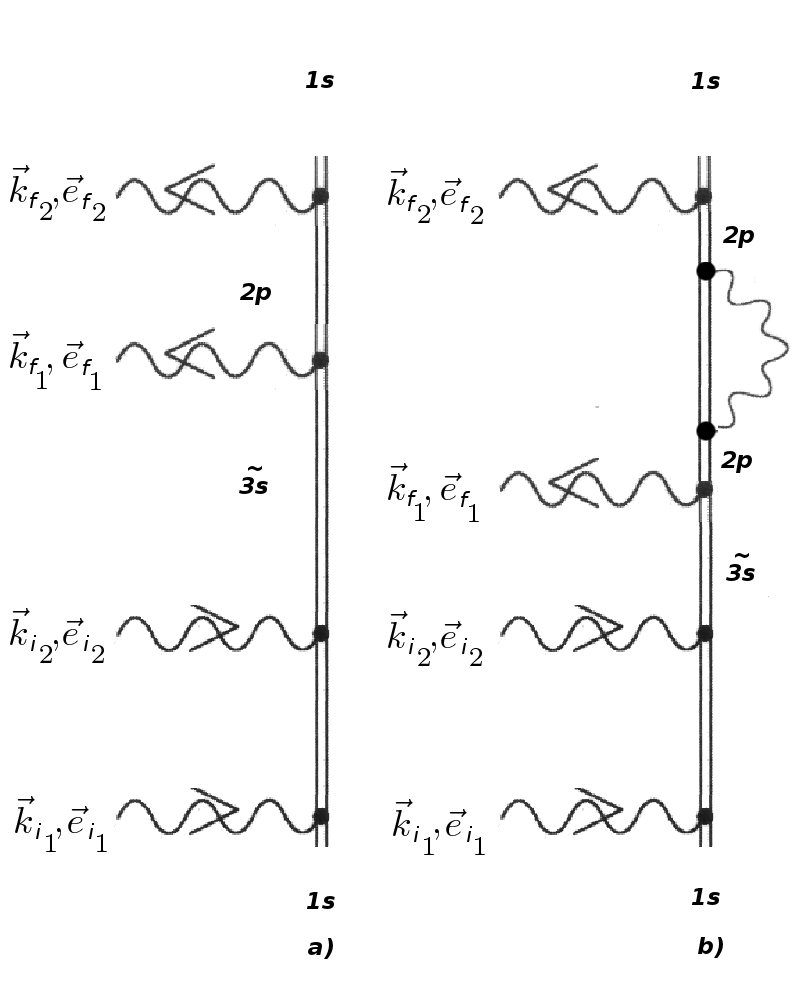}
  \caption{Feynman graph describing the two-photon resonance scattering on the ground state of hydrogen atom, with the excitation of $ 3s $ state (resonance condition $ \omega_{i_1}+\omega_{i_2}=E_{3s}-E_{1s} $) and the decay cascade resonances $ \omega^{res.1}=E_{3s}-E_{2p} $, $ \omega^{res.2}=E_{2p}-E_{1s} $. In FIG. 3$ a $ the basic process of the resonant scattering with the excitation of the $ 3s $ level and decay $ 3s-2p-1s $ is depicted. In FIG. 3$ b $ the electron self-energy insertion in the upper electron propagator is made. Notation $ \tilde{3s} $ means that the Low procedure is already performed for this electron line. The other notations are the same as FIGS. 1, 2}
\end{figure}
\begin{figure}[h!]
  \centering
\includegraphics[scale=0.4]{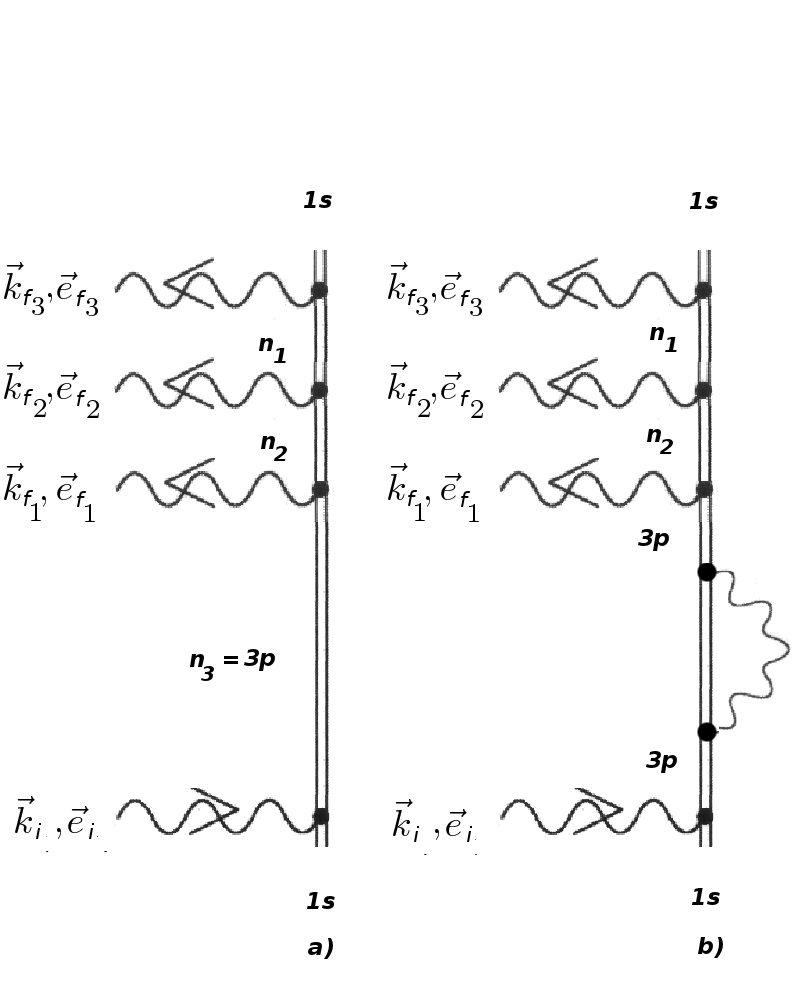}
  \caption{Feynman graph describing the resonant process with one-photon absorption and three-photon emission for the ground state of the hydrogen atom. In FIG. 4$ a $ the basic process is shown and in FIG. 4$ b $ the electron self-energy insertion in the central electron propagator is made. The notations are the same as in FIG. 1}
\end{figure}
\end{document}